# Large-Aperture All-Solid-State Cascaded Liquid-Crystal Beam Steering for High-Resolution Wide-Field Imaging

Chenxi Liu[1], Yongxiang Qu[1], Xiaoxin Wang[2], Leixin Meng[1,2], Xiaohua Feng[1,2], Qidong Wang[3], Guohao Zhang[1], Xuejun Zhang[1,2], Yubing Han[1], Qing Yang[1,2], Xu Liu[1,2], Mingwei Tang[1,2,*] and Kai Wei[1,2,*]

[1] State Key Laboratory of Extreme Photonics and Instrumentation, Zhejiang Key Laboratory of Autonomous Optoelectronic Perception, College of Optical Science and Engineering, Zhejiang University, Hangzhou 310027, China.

[2] ZJU-Hangzhou Global Scientific and Technological Innovation Center, Zhejiang University, Hangzhou 311215, China.

[3] Changchun Institute of Optics, Fine Mechanics and Physics, Chinese Academy of Sciences, Changchun 130033, China

* Address all correspondence to:

Mingwei Tang (tangmw@zju.edu.cn); Kai Wei(kwei@zju.edu.cn);

**Abstract**: High-resolution wide-field imaging is essential for applications requiring simultaneous global coverage and local detail, yet conventional approaches face a fundamental trade-off: wide-FOV cameras sacrifice spatial sampling density by distributing finite detector pixels over a broad angular range, while telephoto systems resolve fine features at the cost of scene coverage. Beam-steering devices can mitigate this trade-off but is currently limited to achieve simultaneously all-solid-state, large aperture, and high speed. Here, we report an all-solid-state large-aperture cascaded liquid-crystal beam-steering (CaLiBS) imaging system that extends the effective angular range of a high-resolution narrow-FOV camera by electrically steering sub-FOVs. The CaLiBS module comprises cascaded liquid crystal waveplates and liquid crystal Pancharatnam-Berry phase gratings; a theoretical voltage-prediction model with a hierarchical search algorithm enables efficient calibration under oblique incidence and 10 times faster calibration speed compared with conventional method. The calibrated system addresses sub-FOVs across ±15.15° × ±15.15°at 2° intervals with diffraction efficiency above 60%. Sequential sub-FOV acquisition reconstructs a 34.7° × 34.7° composite image—an 8.6-fold enhancement in spatial–bandwidth product over a single-shot wide-FOV camera using the same detector. Combined with object tracking methods, sub-FOV switching further enables high-resolution tracking of moving vehicles within the wide-area scene. This cascaded LC architecture offers a scalable pathway toward compact, vibration-free, and high-resolution wide-field observation.

# 1 Introduction

High-resolution wide-field optical imaging is indispensable across diverse domains, including biomedical imaging[1–3], traffic monitoring[4], remote sensing[5], and security surveillance[6,7]. In these scenarios, an imaging system must simultaneously cover a large angular field, and preserve sufficient local spatial sampling to resolve fine target features within regions of interest. Wide-area surveillance, for instance, demands global situational awareness, yet reliable target identification requires detailed local information such as object morphology. Conventional single-sensor optical systems, however, are fundamentally constrained by a trade-off between field of view (FOV) and spatial resolution. Governed by etendue conservation, wide-FOV optics typically require an aperture stop substantially smaller than the lens clear aperture [8]. Expanding the FOV therefore disperses the finite detector pixels and optical bandwidth over a broader angular range, degrading local resolution and obscuring fine structural detail. For an imaging system, the spatial bandwidth product (SBP), defined as the product of spatial extent and spatial frequency bandwidth, can be written as[9,10]:

$$SBP \propto A \times f^2 = \frac{A}{\delta^2} = \frac{A}{(\lambda/NA)^2 + {\delta_g}^2} \tag{1}$$

where $A$ denotes the image range associated with the FOV, $\delta$ is the diameter of the diffuse spot determined by the *NA* and the geometrical aberration $\delta_g$. Since the total SBP is fundamentally limited by the optical aperture, the spatial extent and spatial frequency bandwidth compete for the same bandwidth resource, making it difficult to simultaneously achieve both large FOV and high spatial resolution within a single static imaging channel.

To overcome this bottleneck, two mainstream solutions have been proposed: multi-device array imaging[1] and single-device scanning imaging[11]. Multi-device array imaging achieves large-FOV coverage by integrating multiple high-resolution lenses or sensors and stitching images from different perspectives. However, the integration of multiple cameras inevitably leads to bulky system configurations, complex inter-device calibration requirements[12], and increased data transmission and synchronization burdens. These drawbacks limit its applicability in compact or weight-sensitive scenarios, such as micro-endoscopy[13] and vehicle-mounted systems[14].

Single-device scanning imaging systems typically rely on a single high-resolution camera paired with mechanical scanning mechanisms, such as rotation stages or linear translation stages. However, such mechanical motion introduces inherent limitations. The large inertia of mechanical drives leads to overshoot, prolonged settling times, and limited positioning precision. Motion-induced vibrations[14] degrade image quality and introduce artifacts[15], while slow response times preclude dynamic scene imaging. Furthermore, prolonged operation diminishes reliability and increases maintenance overhead. Semi-solid-state scanning systems employ devices such as Risley prisms[16], galvanometer scanners[17], or MEMS scanners[18–20] to deflect beams via reflection or refraction, sequentially directing light from different fields of view onto a stationary camera. Although these approaches reduce system complexity and improve scanning speed through lighter moving components, they remain fundamentally reliant on mechanical motion and are thus still constrained by inertia and dynamic stability limitations.

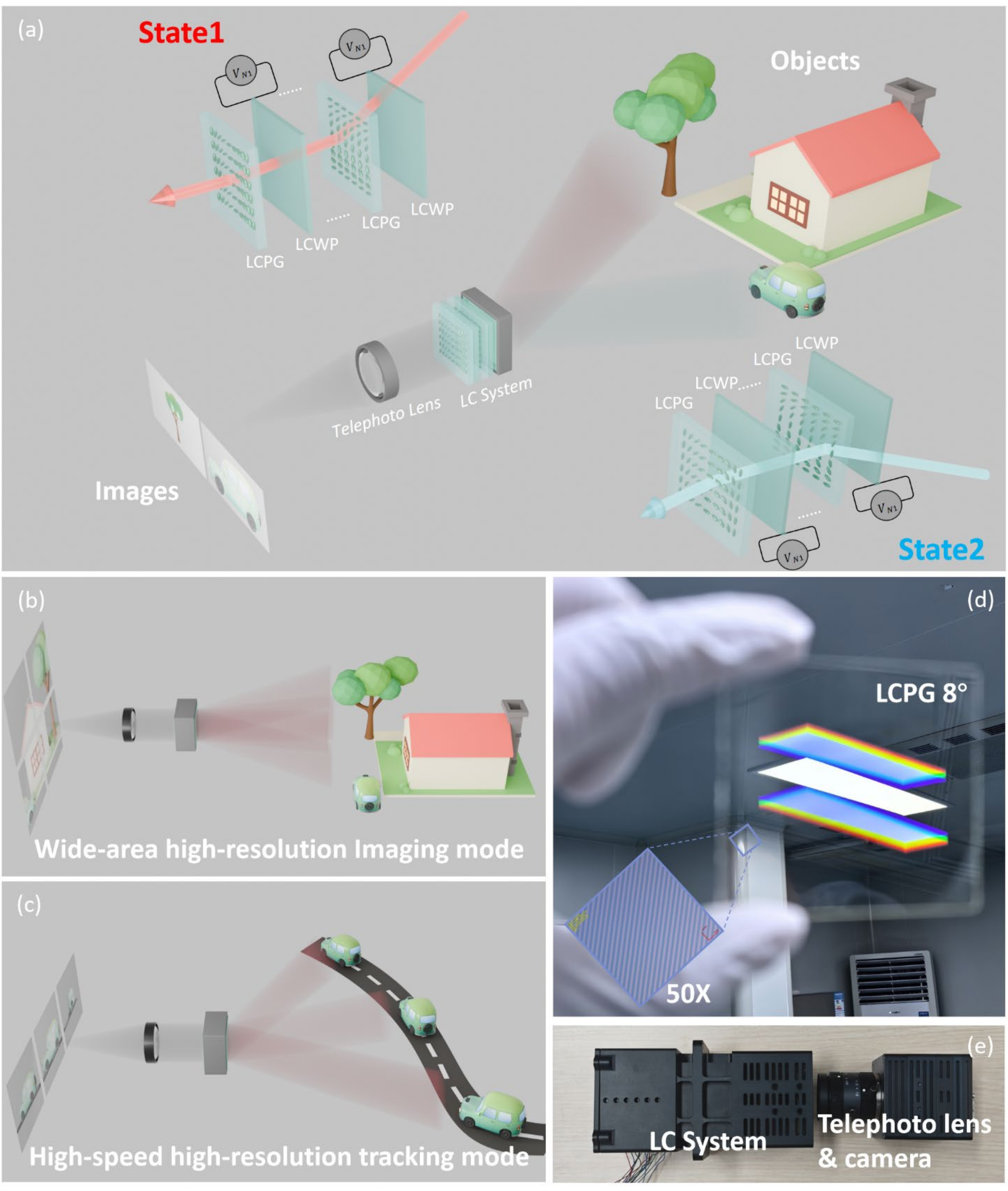

**Fig. 1 | Concept and implementation of the all-solid-state Cascaded Liquid-crystal Beam Steering (CaLiBS) imaging system.** (a) Schematic of the imaging system, which consists of a CaLiBS module and a small-FOV high-resolution telephoto imaging module. By applying different voltage configurations to the LCWP–LCPBG pairs, off-axis light from different sub-FOVs can be redirected toward the near optical axis and captured by the telephoto imaging module with high spatial resolution. Two representative voltage-controlled steering states, State 1 and State 2, are shown to illustrate how the cascaded LC system selects different sub-FOVs without mechanical motion. (b) Wide-field high-resolution (WFHR) mode, in which the CaLiBS sequentially switches among

different steering states to scan multiple sub-FOVs and reconstruct an enlarged FOV. (c) High-speed high-resolution (HSHR) tracking mode, in which the CaLiBS system redirects the selected sub-FOV according to the target position, enabling high-resolution imaging of moving objects within a large field of regard. (d) Photograph of a representative LCPBG device, showing an 8° LCPBG element and its polarized optical microscopic pattern. (e) Photograph of the experimental prototype, including the CaLiBS system, a telephoto lens, and a camera.

To further improve integration, response speed, and stability, various all-solid-state beam steering approaches have been developed, including electro-optic crystal scanners[21,22], silicon-based optical phased arrays (Si-OPAs)[23–25], acousto-optic modulators (AOMs)[26,27], and liquid crystal (LC) devices[28,29]. Electro-optic crystals offer ultrafast response but require high driving voltages and face challenges in scaling to large apertures. Si-OPAs achieve beam steering through phase modulation of array elements with high angular precision; however, complex optical addressing circuitry reduces the aperture fill factor, diminishing beam steering efficiency. AOM-based steering enables microsecond-level response but is constrained by the acoustic interaction length and diffraction efficiency, with cascaded configurations introducing considerable optical loss. Although polarization-multiplexed metasurfaces cascaded with LC layers have been proposed to extend the FOV, the fabrication-limited small aperture of metasurfaces restricts their practical deployment[28]. Collectively, these all-solid-state approaches are fundamentally constrained in optical aperture, which limits their applicability to high-resolution wide-field imaging.

Here we propose a high-resolution wide-field imaging method with cascaded beam steering systems based on liquid crystal wave plate (LCWP) and liquid crystal Pancharatnam-Berry phase gratings (LCPBG), which exhibit distinctive advantages in efficiency and aperture utilization. In LCPBG, the phase profiles desired for grating are realized with the Pancharatnam–Berry(PB) phases by using a patterned LC half-wave plate. The orientation of LC molecules at each spatial point on the grating surface determines the local phase change, which is similar to the orientation of the unit element in conventional metasurfaces[30]. The LCWP-LCPBG cascaded architecture eliminates the need for pixelated addressing circuits, enabling an aperture ratio approaching 100%. LC devices can be fabricated into large-aperture components at relatively low cost—a feat that remains challenging for other all-solid-state methods. Furthermore, it features minimal wavefront distortion, with the diffraction efficiency of a single LCPBG capable of approaching 100% under ideal conditions[31,32]. Despite the widespread adoption of cascaded LCPBGs in beam steering applications such as free-space optical (FSO) communication systems[33], light detection and ranging (LiDAR)[34,35], optical display systems[36], and integrated multifunctional computational imaging platforms[37], systematic investigations into their application in optical imaging remain scarce.

Moreover, most existing research remains limited to one-dimensional spot array switching based on a small number of cascaded layers[38]. As the number of cascaded LCPBG layers increases, efficiency optimization becomes considerably more challenging, and the required voltage calibration procedure becomes prohibitively time-consuming, severely impeding the development of large-scale cascaded systems[39]. Consequently, there remains a lack of comprehensive investigation and efficient calibration methods for large-scale cascaded LCPBG-based two-dimensional (2D) beam steering devices, particularly for applications in optical imaging systems that demand high efficiency and low crosstalk. These challenges underscore the urgent need to develop high-precision, high-speed voltage calibration strategies for large-scale cascaded LCPBG systems, with the goal of substantially reducing calibration time, improve voltage tuning accuracy, and push overall system efficiency toward its theoretical limit.

In this work, we present an all-solid-state Cascaded Liquid-crystal Beam Steering (CaLiBS) imaging system that provides a cost-effective way to overcome the traditional trade-off between FOV, resolution, and imaging speed, as illustrated in **Fig.1**(a). The system integrates a compact CaLiBS module——comprising large-aperture cascaded LCWPs and LCPBGs——with a narrow-FOV, high-resolution camera. This architecture supports two task-dependent operation modes: wide-field high-resolution (WFHR) mode for static or slowly varying scenes (**Fig.1**(b)), and high-speed high-resolution (HSHR) tracking mode for moving targets in dynamic wide-area scenes **Fig.1**(c). The fabricated LCWPs and LCPBGs feature a 50 mm×50 mm aperture(**Fig. 1**(d)), high efficiency and excellent wavefront quality, enabling high-quality imaging performance. The prototype is shown in **Fig.1**(e). A theoretical prediction model for LCWP driving voltages, combined with a hierarchical search-based optimization strategy, has been developed to achieve efficient and rapid calibration of the cascaded steering states under oblique incidence. With the calibrated CaLiBS module, different sub-FOVs can be electrically directed into the same high-resolution imaging channel with high efficiencies and diminished crosstalk. Experiments verify the advantages of the CaLiBS in high-fidelity reconstruction of real indoor and outdoor scene images, and dynamic tracking imaging and precise recognition of targets of interest. The proposed CaLiBS provides an all-solid-state route for enhancing the wide-field imaging capability of high-resolution optical systems dismissing mechanical scanning and has important application value in fields such as biomedical imaging, semiconductor inspection, autonomous driving, remote sensing, and urban traffic monitoring.

# 2 Results

### 2.1 System overview of CaLiBS and the principle of SBP enhancement

The CaLiBS module is constructed from cascaded LCWP–LCPBG steering units. Based on the polarization-selective ±1st-order diffraction of LCPBGs, each LCWP–LCPBG unit functions as a binary angular steering element, in which the LCWP controls the incident polarization handedness and the LCPBG converts the circular polarization light into the designed diffraction angle. Detailed operation principles of a single LCWP–LCPBG unit are provided in **Supplementary Note 1**.

In the cascaded configuration, the total steering angle is determined by the diffraction orders selected by all active units. For a two-dimensional CaLiBS module, we define two units of cascaded LCWPs and LCPBGs which deflect beams in the x-z and y-z planes respectively as a layer. In this module, the cascaded momenta would be the summation of momenta of each single pattern.

$$\sum_{i=1}^{N} \varphi\left(x, y, k_{x,i}, k_{y,i}\right) = \sum_{i=1}^{N} \left(k_{x,i}x + k_{y,i}y\right) \tag{2}$$

Thus, the output steering angle can be expressed as:

$$sin\theta_x = \sum_{i=1}^{N} S_{xi} \times sin\alpha_{xi} \quad (3)$$

$$sin\theta_y = \sum_{i=1}^{N} S_{yi} \times sin\alpha_{yi} \quad (4)$$

where $\theta_x, \theta_y$ are deflected angles in x-z plane and y-z plane; $S_{xi}, S_{yi}$ represent diffraction orders of the LCPBGs in the i-th layer, taking value of $\pm 1$; $\alpha_{xi}, \alpha_{yi}$ are diffraction angles of the i-th layer's LCPBGs; $N$ represents the number of the LCWP-LCBPG units in x-z or y-z planes. In our implementation, LCPBGs with diffraction angles of 1°, 2°, 4°, and 8° are cascaded in each direction, generating $2^4 = 16$ steering states from about −15.15° to +15.15° with a 2° interval. The complete two-dimensional CaLiBS module therefore provides a $2^N \times 2^N = 16 \times 16$ steering grid over a 30° × 30° angular range. Equations (3) and (4) indicate that for an expecting deflecting angle $(\theta_x, \theta_y)$, diffraction order of each LCPBG can be uniquely determined, which also means that the propagation path in the cascaded system is determined. The known light propagation paths help determine the incident angle of the light on each LCWP. The diffraction order values and their corresponding deflection angles are provided in **Table S2** of **Supplementary information**. The LCPBGs are fabricated using polarization holography lithography based on photoalignment. Details of the fabrication process can be found in **Methods** section. **Fig.1(**d**)** shows the diffraction photographs and polarized optical microscopic images of the fabricated exemplified 8° LCPBGs at 1550 nm. The grating areas exhibit good uniformity without observable macroscopic defects.

To rigorously control the influence of CaLiBS on imaging quality and avoid introducing severe aberrations, we have carefully considered key performance parameters of the LC devices during fabrication, including efficiency and wavefront-related metrics. Specifically, the intrinsic diffraction efficiency—defined experimentally as the ratio of the optical power diffracted into the target order (e.g., +1st order) to the total incident optical power, measured under a specific polarization state and wavelength —is greater than 99.5%, indicating highly uniform LC director alignment and excellent grating quality. The Fresnel loss arising from reflections at interfaces with dissimilar refractive indices is below 0.1%, while the absorption loss attributed to the transparent-conducting-electrode layers is less than 0.2%. Additionally, the scattering loss caused by random light scattering from the LCPBGs away from the desired steering direction is controlled below 0.3%. The wavefront of a single device is critical for optical imaging applications, especially for large-scale cascaded systems. To reduce the wavefront error, we have carefully controlled the substrate of LCPBG and LCWP, as well as LC fabrication processes. Analysis of wavefront and point spread function (PSF) of cascaded system is shown in **Supplementary Note 2**, with root-mean-square (RMS) errors of 1/20 $\lambda$ and 1/9 $\lambda$ for single LCWP and single LCPBG at the measurement wavelength of 632.8 nm.

**Large SBP imaging with CaLiBS**

In conventional imaging systems, a fixed detector and optical aperture must cover the entire angular field simultaneously. As a result, the local transferable spatial-frequency bandwidth becomes field dependent and typically decreases as the FOV is enlarged, due to reduced local sampling density and increased off-axis aberrations. The effective SBP can be expressed as the integrated accumulation of local transferable frequency-domain support over the entire field:

$$SBP_{conv} \propto \int_{\Omega} |B(\boldsymbol{\theta})| d\boldsymbol{\theta} \tag{5}$$

where $\Omega$ denotes the angular of the wide-FOV system, $\theta$ is the field-angle coordinate, and $B(\theta)$ represents the usable local spatial-frequency support at that field point. This expression indicates that simply increasing the geometric FOV does not necessarily produce a proportional increase in effective SBP, because the usable local bandwidth $B(\theta)$ may degrade across the field.

The proposed CaLiBS imaging system follows a different strategy. Instead of capturing the entire wide field with a single short-focal-length wide-FOV lens, CaLiBS partitions the total angular field into multiple sub-FOVs and sequentially steers them into the same high-resolution narrow-FOV imaging channel. In this configuration, different sub-FOVs share approximately the same high-quality imaging channel, and the local spatial-frequency support of the high-resolution channel is reused across the scanned field. Therefore, the effective SBP of CaLiBS can be expressed as:

$$SBP_{scan} \propto \sum_{m=1}^{M} \int_{\Omega_m} |B_0| d\theta = |B_0| \sum_{m=1}^{M} |\Omega_m| \tag{6}$$

where $\Omega_m$ is the m-th steered sub-FOV, $M$ is the number of scanned sub-FOVs, and $B_0$ denotes the usable spatial-frequency support of the high-resolution narrow-FOV imaging channel.

When the sub-FOVs are approximately uniform in size, *i.e.*, $|\Omega_{\mathrm{m}}| \approx |\Omega_0|$, this simplifies to:

$$SBP_{scan} \propto M|\Omega_0||B_0| \tag{7}$$

Unlike a conventional wide-FOV camera, CaLiBS does not rely on a single optical channel to simultaneously maintain high spatial-frequency transfer over the entire field. Instead, it extends the accessible angular coverage by capturing multiple channels while preserving the local imaging capability of the high-resolution channel. A detailed derivation of the SBP model is provided in **Supplementary Note 3**.

### 2.2 Hierarchical search strategy for efficiency optimization

A key practical challenge in the CaLiBS system is efficiency degradation induced by oblique incidence. In a cascaded LCWP–LCPBG architecture, the incident angle on each LCWP depends on the diffraction orders selected by the preceding LCPBGs rather than being fixed. As illustrated in **Fig. 2**(a), the phase retardation of an LCWP depends on the LC material parameters, the beam propagation direction, and the relative orientation between the incident polarization and the LC director. Under oblique incidence, the effective optical path and birefringence vary with incident angle, rendering the half-wave and full-wave driving voltages angle-dependent. Driving the LCWPs with voltages calibrated solely under normal incidence accumulates retardation errors, leading to incomplete polarization conversion and reduced diffraction efficiency in the cascaded output.

To address this, we established an angle-dependent voltage prediction model for the LCWPs (see **Supplementary Note 4**). The model computes the half-wave and full-wave driving voltages as functions of incident angle by accounting for the effective optical path and angle-dependent birefringence of the LC layer. The calculated voltage maps, shown in **Fig. 2**(b,c), provide theoretical predictions over the incident-angle range covered by the system. These maps serve as initial voltage configurations for subsequent cascaded-system calibration. **Fig. 2**(d,e) shows the measured half-wave and full-wave voltages at different incident angles, demonstrating excellent agreement with the calculated values.

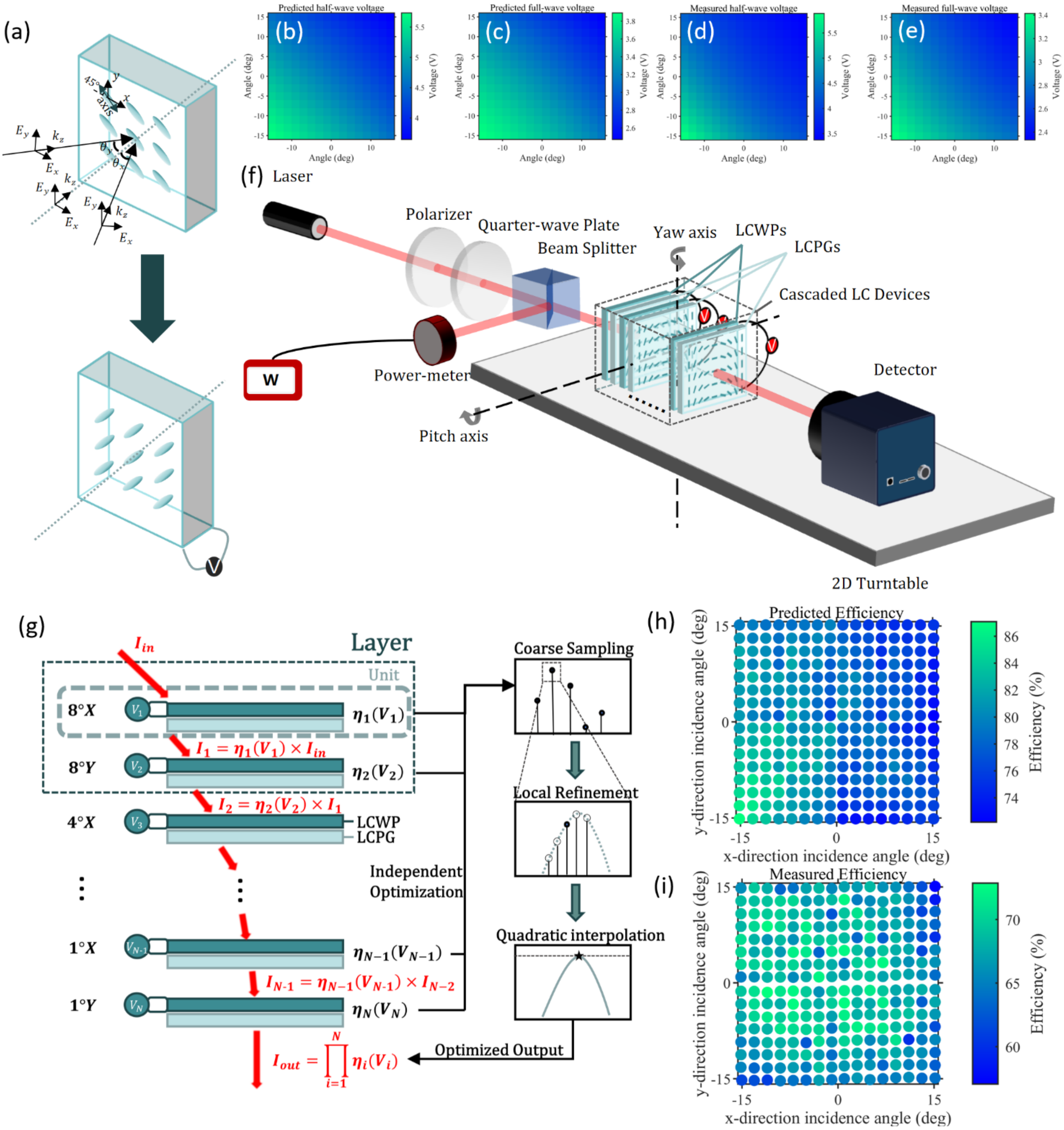

**Fig. 2 | Experimental calibration of the CaLiBS system.** (a) Diagram showing the coordinate system and the orientation of the LC director axis (at 45° to the x-axis) within an LCWP. The incident beam at different angles can be decomposed into a latitudinal component $E_x$ and a longitudinal component $E_y$, where $k_z$ represents the wave vector of the incident beam, and $\theta_x$ and $\theta_y$ denote the latitudinal and longitudinal angles of incident beam. (b-e) Comparison of calculated (b, c) and experimentally measured (d, e) driving voltages required for the LCWP to achieve half-wave (π) and full-wave (2π) phase modulation as a function of the incident angle. (f) Optical setup for

characterizing the beam steering performance. A 1550 nm laser is converted into circularly polarized light using a polarizer and a quarter-wave plate. A beam splitter then directs 50% of the light to a reference power meter, while the remaining beam passes through the LC cascaded system. The system and detectors are mounted on a 2D turntable to measure deflection angles and efficiencies. (g) Photograph of the fully assembled 4-layer LC beam steering system and optimization process, comprising four LCWP-LCPBG units for each of the x- and y-scanning directions, enabling a $2^4 \times 2^4$ steering grid. The device deflects obliquely incident beams within certain FOV to small exit angles, while independent 1D optimizations consists of coarse sampling, local refinement, and quadratic interpolation are applied to each channel for maximum efficiency at this FOV. (h, i) Theoretical and measured 2D efficiency distributions demonstrating an over 30°×30° FOV with an approximately 2° angular resolution. The colormap indicates the diffraction efficiency at each steering angle.

The final output efficiency of CaLiBS is governed by the combined efficiency of all cascaded LCWP–LCPBG units. For a selected steering path, the system efficiency can be expressed as:

$$P(V) = P_{in} \prod_{i=1}^{N} \eta_i (V_i) \tag{8}$$

where $P_{in}$ is the input power, $V_i$ denotes the driving voltage of the i-th LCWP, and $\eta_i$ represents the polarization conversion and diffraction efficiency of that unit. This expression reveals that, for a given steering path, each layer's contribution to the final efficiency can be approximately treated as independent. This separability provides the physical basis for optimizing each LCWP voltage independently rather than performing a full high-dimensional search.

We leverage this separable structure to transform the original high-dimensional optimization problem into a channel-wise one-dimensional peak-searching problem, adopting a derivative-free hierarchical search strategy. As shown in **Fig. 2**(g), the strategy comprises three steps: coarse sampling localization, local refinement search, and quadratic interpolation correction. First, coarse samples within each channel's allowable voltage range rapidly determine the approximate peak location. Subsequently, a local fine search with smaller voltage steps improves peak localization accuracy. Finally, quadratic curve fitting using adjacent sampling points estimates the peak voltage via parabolic interpolation. Since the polarization conversion efficiency of the LCWP exhibits smooth variation with a distinct unimodal characteristic as a function of driving voltage, this hierarchical strategy accurately locates the optimal operating point with minimal sampling.

To calibrate and characterize the CaLiBS system, an optical measurement setup was established (**Fig. 2**(f)). A 1550 nm laser beam, converted to circularly polarized light, is split into a reference arm for power monitoring and a signal arm directed through the CaLiBS system. The detectors and CaLiBS module are mounted on a 2D rotation stage, enabling characterization at different incident angles. A camera and power meter measure the deflection angle and output power, respectively.

In the 4-layer device, the LCPBG diffraction angles are chosen as 1°, 2°, 4°, and 8° in both the x-z and y-z planes, yielding 8 independent voltage channels. For each channel, coarse sampling within the 0–6 V range is followed by local refinement with 0.1 V steps around the optimal coarse point, and quadratic interpolation yields the final peak estimate. The voltage calibration for a single target diffraction direction is completed by sequentially optimizing all channels.

To evaluate the efficiency advantage, we compare this method with channel-wise full-range voltage scanning. A traditional exhaustive strategy scanning 0–6 V with 0.05 V steps requires approximately 121 evaluations per channel; for 8 channels and 256 diffraction positions, the total reaches approximately $2.5 \times 10^5$ evaluations, corresponding to a calibration time of approximately 15 hours. In contrast, the hierarchical search strategy reduces single-channel evaluations to approximately 10–15, decreasing total evaluations per target point to 80–100 and reducing system calibration time from 15 hours to 1.5 hours—a nearly tenfold speedup—while maintaining optimized output power essentially consistent with full-range scanning.

These results demonstrate that for CaLiBS systems with physically separable structures, the optimization strategy lies not in adopting complex high-dimensional algorithms, but in fully exploiting the system's structural features to decompose the multidimensional problem into multiple independent one-dimensional searches. This approach significantly reduces experimental burden while preserving optimization accuracy.

### 2.3 Wide-field high-resolution reconstruction and SBP enhancement

The imaging system is implemented with a high-resolution camera consisting of a telephoto lens (Computar M5018-VSW) and an InGaAs sensor (640 × 512 pixels, 15 μm pixel pitch), positioned at the rear of the CaLiBS module. Since the polarization state of light scattered by target objects is random and unpredictable, a polarizer is placed at the system input to ensure a uniform polarization state, enabling the subsequent waveplates and polarization gratings to function as designed. Light from objects within a given FOV passes through the polarizer and is deflected by the cascaded LC devices before being captured by the camera. Voltages applied to the LCWPs select different diffraction orders, steering light from specific orientations into the imaging channel.

In the wide-field high-resolution (WFHR) mode, the system raster-scans the calibrated sub-FOVs, followed by sub-image unmixing, stitching, and fusion, as illustrated in **Fig. 3**(a). With optimized LCWP voltages, a high-efficiency sub-image array covering a large FOV is obtained through sequential scanning (**Fig. 3**(c)). In certain sub-images, artifacts inconsistent with the corresponding FOV positions are observed, arising from crosstalk between sub-FOV channels.

Systematic analysis reveals that these artifacts originate predominantly from the diagonally positioned sub-image α', which exhibits a centrosymmetric relationship with sub-image α about either the full-FOV center or a half-FOV center in the sub-image array. This reciprocal crosstalk arises because certain LCWP–LCPBG layers exhibit diffraction efficiencies below 100%, allowing both +1 and −1 diffraction orders to propagate simultaneously, whereas ideally only one order should be transmitted. A detailed analysis of the crosstalk mechanism is provided in **Supplementary Note 8**.

Consequently, the captured intensity represents a linear superposition of the light intensities from diagonal FOVs. This linear crosstalk is modeled as:

$$Y_{\alpha} = I_{\alpha} + aI_{\alpha'} \tag{9}$$

$$Y_{\alpha'} = I_{\alpha'} + bI_{\alpha} \tag{10}$$

where $Y\alpha$ and $Y\alpha'$ denote the captured images, $I\alpha$ and $I\alpha'$ represent the latent clean images, and $a$, $b$ are mixing coefficients characterizing the cross-view interference strength. Assuming $1 - ab \neq 0$, the system admits a closed-form inverse:

$$\begin{cases} I_\alpha = \dfrac{Y_\alpha - aY_{\alpha'}}{1-ab} \\ I_{\alpha'} = \dfrac{Y_{\alpha'} - bY_\alpha}{1-ab} \end{cases} \tag{11}$$

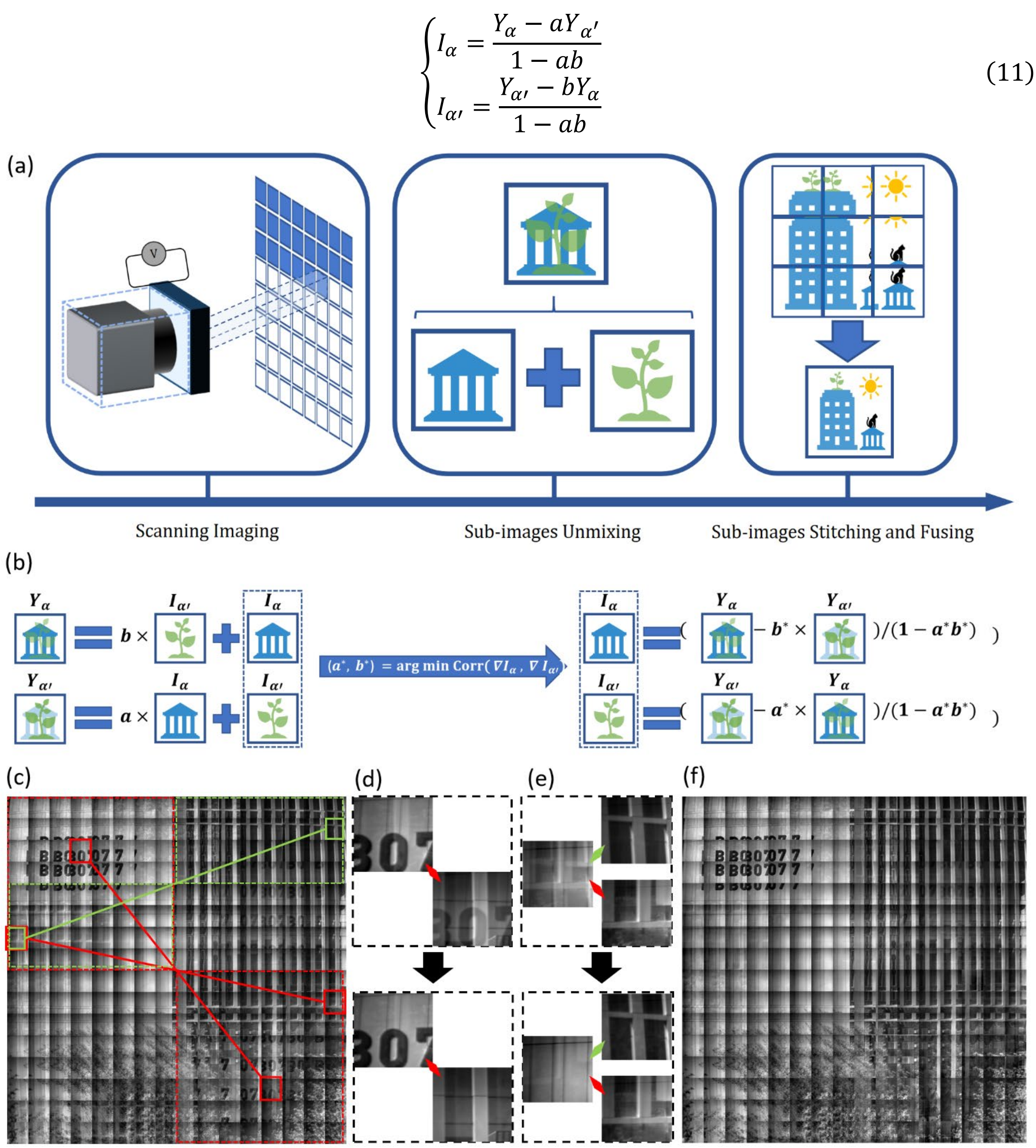

**Fig. 3 | Imaging process and crosstalk calibration of CaLiBS.** (a) Workflow of the scanning imaging system. A high-resolution camera captures sub-images as the LC system steers the FOV. These sub-images are then unmixed, stitched, and fused to reconstruct the final large-FOV image. (b) Schematic representation of the linear crosstalk phenomenon and the unmixing principle. Sub-images from diagonal FOVs ($I_\alpha$ and $I_{\alpha'}$) can contaminate each other due to imperfect diffraction efficiency, resulting in captured images ($Y_\alpha$ and $Y_{\alpha'}$). The mixing coefficients $a$ and $b$ are estimated by minimizing the correlation of the spatial gradients of the unmixed images. (c) Raw sub-image array captured by the system before crosstalk correction. (d, e) Examples of the unmixing process for sub-image pairs exhibiting (d) full-field symmetric crosstalk and (e) more complex, multi-scale crosstalk.

(f) The full sub-image array after applying the linear unmixing model to each individual sub-image. Compared to the raw array in (c), the crosstalk artifacts have been effectively removed.

Since $I\alpha$ and $I\alpha'$ originate from different views, their spatial structures are expected to be statistically independent. Based on this assumption, we estimate the mixing coefficients by minimizing the correlation between the spatial gradients of the unmixed images:

$$\mathcal{L}(a,b) = Corr(\nabla I_{\alpha}, \nabla I_{\alpha'}) \tag{12}$$

where $\nabla$ denotes the image gradient operator and $Corr(\cdot)$ represents a normalized correlation metric. The optimal coefficients are obtained via 2D grid search over feasible ranges of $a$ and $b$, requiring no labeled data or prior training. **Fig.3**(b) illustrates the unmixing workflow.

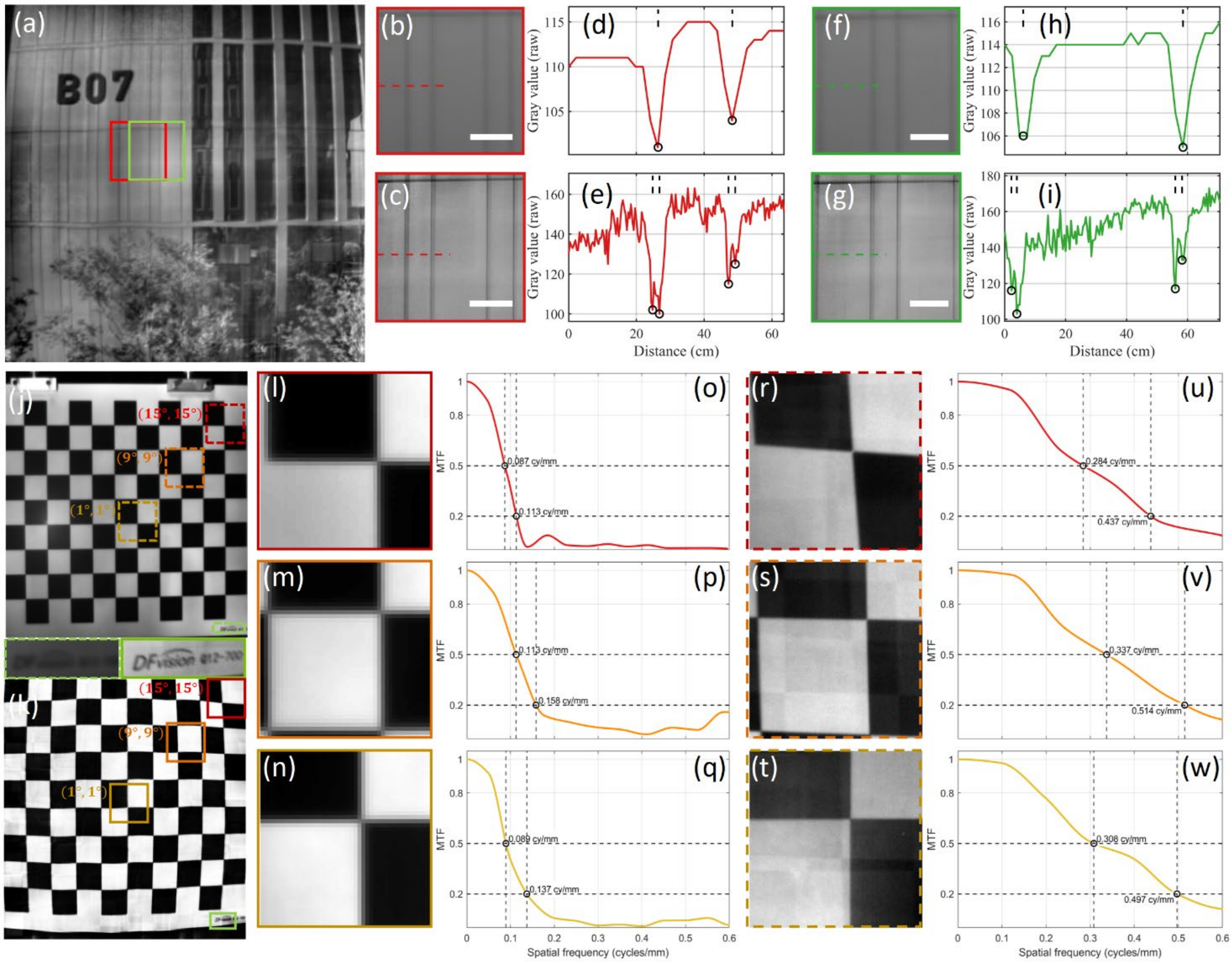

**Fig. 4 | Imaging performance comparison with a commercial large-FOV lens.** (a) The final reconstructed full-FOV image obtained by the CaLiBS system, with the locations of the detailed views indicated by boxes. (b, c) Partial enlarged views corresponding to the red-boxed region in (a), captured by a commercial large-FOV camera and the CaLiBS imaging system, respectively. (f, g) Partial enlarged views corresponding to the green-boxed region in (a), captured by a commercial large-FOV camera and the CaLiBS imaging system, respectively. (d, e, h, i) Intensity profiles along the lines marked in (b, c, f, g), quantitatively demonstrating the superior resolution and contrast of the CaLiBS system. (j, k) Full-FOV images of a checkerboard target captured by the large-FOV

camera (j) and the CaLiBS imaging system (k), respectively. In particular, the image (k) captured by the CaLiBS system is processed by mosaicking and distortion correction. The insets show magnified raw views of the extreme edge FOV, indicated by the dashed green box in (j) and the solid green box in (k). (l-n) Images of the checkerboard from the large-FOV camera at FOVs of (15°,15°), (9°,9°), and (1°,1°). (o-q) Corresponding MTF curves for the images in (l-n). (r-t) Unmixed original sub-images of the checkerboard from the CaLiBS system at FOVs of (15°,15°), (9°,9°), and (1°,1°). (u-w) Corresponding MTF curves for the sub-images in (r-t).

For the experimentally captured sub-images, the unmixing results are presented in **Fig. 3**(d–e). The sub-image pairs highlighted by red dashed boxes exhibit interference dominated by full-field-scale crosstalk centered at the full-FOV symmetric center. In contrast, the pairs marked by green dashed boxes display central symmetry with respect to half-field region centers, indicating secondary interference mechanisms operating at the sub-field scale. As shown in **Fig. 3**(d), the sub-image pair from the red-boxed region in **Fig. 3**(c), dominated by full-field-scale interference, is effectively unmixed using the linear model. The sub-image relationships in **Fig. 3**(e) are more complex, with simultaneous contributions from both full-field and half-field centrosymmetric components; nevertheless, the linear unmixing model still efficiently eliminates the artifacts.

**Fig.3**(f) displays the complete sub-image array after applying unmixing to each sub-image. Most prominent artifacts—such as the "B07" marking from a different FOV appearing on the building facade—are successfully removed.

To demonstrate the capability of simultaneously achieving large-FOV and high-resolution optical imaging, we compare the performance of our CaLiBS system with a commercial large-FOV camera, which uses the same imaging sensor but is equipped with a large-FOV lens (Computar M0814-VSW). **Fig. 4**(a) illustrates the full-FOV imaging result obtained by stitching and fusing all sub-images captured by the CaLiBS system, with the positions of the detailed views in **Fig. 4**(b, c, f, g) indicated by red and green boxes. **Fig. 4**(b, c) and **Fig. 4**(f, g) present partial views of the building façade acquired by the commercial large-FOV camera and the CaLiBS system, respectively. In the experimental scene, each vertical line is composed of two finer parallel lines. As shown in **Fig. 4**(f, g), the CaLiBS system successfully resolves each line pair, whereas the commercial large-FOV camera presents blurred line structures, as shown in **Fig. 4**(b, c). For quantitative comparison, the intensity profiles along the dashed lines in these images are plotted in **Fig. 4**(d, e, h, i). The profiles from the CaLiBS images in **Fig. 4**(h, i) show two clearly distinguishable valleys for each line pair, whereas the profiles from the large-FOV camera images in **Fig. 4**(d, e) are smoother and contain only weak or unresolved modulation. These results demonstrate the superior local spatial resolution and contrast of the CaLiBS system over a large-FOV camera.

To further quantitatively compare the SBP improvement of the CaLiBS system, we also performed an imaging experiment using a checkerboard target. **Fig. 4** (j) and **Fig. 4**(k) present the images directly captured by the large-FOV camera and the CaLiBS system, respectively. Positioned between these two full images are magnified observation regions of local features, indicated by green dashed and solid boxes: the left panel with the dashed box corresponds to the character information area in the bottom-right edge FOV of **Fig. 4**(j), while the right panel with the solid box corresponds to the equivalent edge-FOV position in **Fig. 4**(k). Comparative evaluation of the edge-FOV performance between the two imaging datasets reveals that the character information within the edge FOV from the CaLiBS system (**Fig. 4**(k)) retains excellent clarity and legibility. In stark contrast, the characters at the identical edge-FOV position from the large-FOV camera (**Fig. 4**(j)) exhibit substantial degradation of detail, with legibility severely compromised. This pronounced discrepancy directly demonstrates the superior resolution performance of the CaLiBS system with high-resolution camera at the edge of the large FOV.

Furthermore, the modulation transfer function (MTF) was calculated at multiple FOV positions for each system. **Fig. 4**(l)–4(n) show sub-images captured by the large-FOV camera at FOV positions of (15°, 15°), (9°, 9°), and (1°, 1°), respectively, while the corresponding images acquired by the CaLiBS system are displayed in **Fig. 4**(r)–4(t). For each field position, the MTF curves of both systems are compared side by side in **Fig. 4**(o)–4(q) and **Fig. 4**(u)–4(w).

The large-FOV camera (**Fig. 4**(l)–4(n)) exhibits pronounced edge blurring and reduced contrast at high spatial frequencies. Its MTF curves (**Fig. 4**(o)–4(q)) show a steep initial roll-off, with the 50% MTF cutoff at approximately 0.087, 0.113, and 0.089 cycles/mm, and the 20% MTF cutoff near 0.113, 0.158, and 0.137 cycles/mm at the three FOV positions, respectively. In contrast, the CaLiBS system (**Fig. 4**(r)–4(t)) preserves sharp edge transitions and maintains smooth intensity gradients across the checkerboard pattern. Its MTF curves (**Fig. 4**(u)–4(w)) decay gradually with increasing spatial frequency, achieving 50% MTF cutoff frequencies of approximately 0.285, 0.337, and 0.308 cycles/mm, and 20% MTF cutoff frequencies near 0.437, 0.514, and 0.497 cycles/mm at the corresponding FOV positions. These results demonstrate that the CaLiBS system provides superior spatial resolution, improved contrast retention, and more robust off-axis performance, as confirmed by both the higher MTF cutoff frequencies and the visibly sharper edges in the captured images.

The SBP was further computed from the measured FOV and $MTF_{50}$ for both systems. The large-FOV camera achieves a 28.3° × 28.3° FOV with an average $MTF_{50}$ of 0.096 cycles/mm, yielding $SBP_{2D} = (W \times f_{50})^2 \approx 2.8 \times 10^3$. The CaLiBS system, with a 34.7° × 34.7° FOV and an average $MTF_{50}$ of 0.310 cycles/mm, achieves $SBP_{2D} \approx 2.4 \times 10^4$, representing an 8.6-fold improvement.

### 2.4 High-speed tracking and high-resolution recognition of moving vehicles

The CaLiBS system can also operate in a selective sub-FOV mode for dynamic wide-field scenes. Instead of scanning all sub-FOVs to reconstruct the full field, the system switches only to the sub-FOVs containing the target, thereby allocating the high-resolution instantaneous FOV to the region where local detail is required.

To demonstrate this capability, we deployed the CaLiBS imaging system at an intersection and recorded a vehicle moving along representative trajectories, as shown in **Fig. 5**(a). The panoramic image was captured by the commercial wide-FOV camera from a fixed viewpoint and was used to provide global scene context. The red curves and arrows indicate the vehicle trajectories and traveling directions, while the yellow markers S1–S5 denote selected sequence positions along the trajectories. The right-turn trajectory was sampled at S1, S2, and S3, whereas the straight trajectory was sampled at S1, S4, and S5, with S1 defined as the initial observation point at t=0 s for each trajectory. The target-driven tracking and switching logic is illustrated in **Fig. 5**(b). After the vehicle is localized in consecutive frames, the displacement of its image-plane position over a short time interval is used to estimate its motion direction and velocity. The next target position is then predicted by extrapolating this motion over the following frame interval. If the predicted position remains within the predefined horizontal and vertical switching boundaries, the current steering state is retained and the localization–prediction cycle is repeated. Once the predicted position crosses one or two boundaries, the controller identifies the corresponding neighboring sub-FOV among the eight possible switching directions and loads the pre-calibrated LCWP voltage set. The target is then relocalized in the newly acquired frame, closing the tracking-and-switching loop.

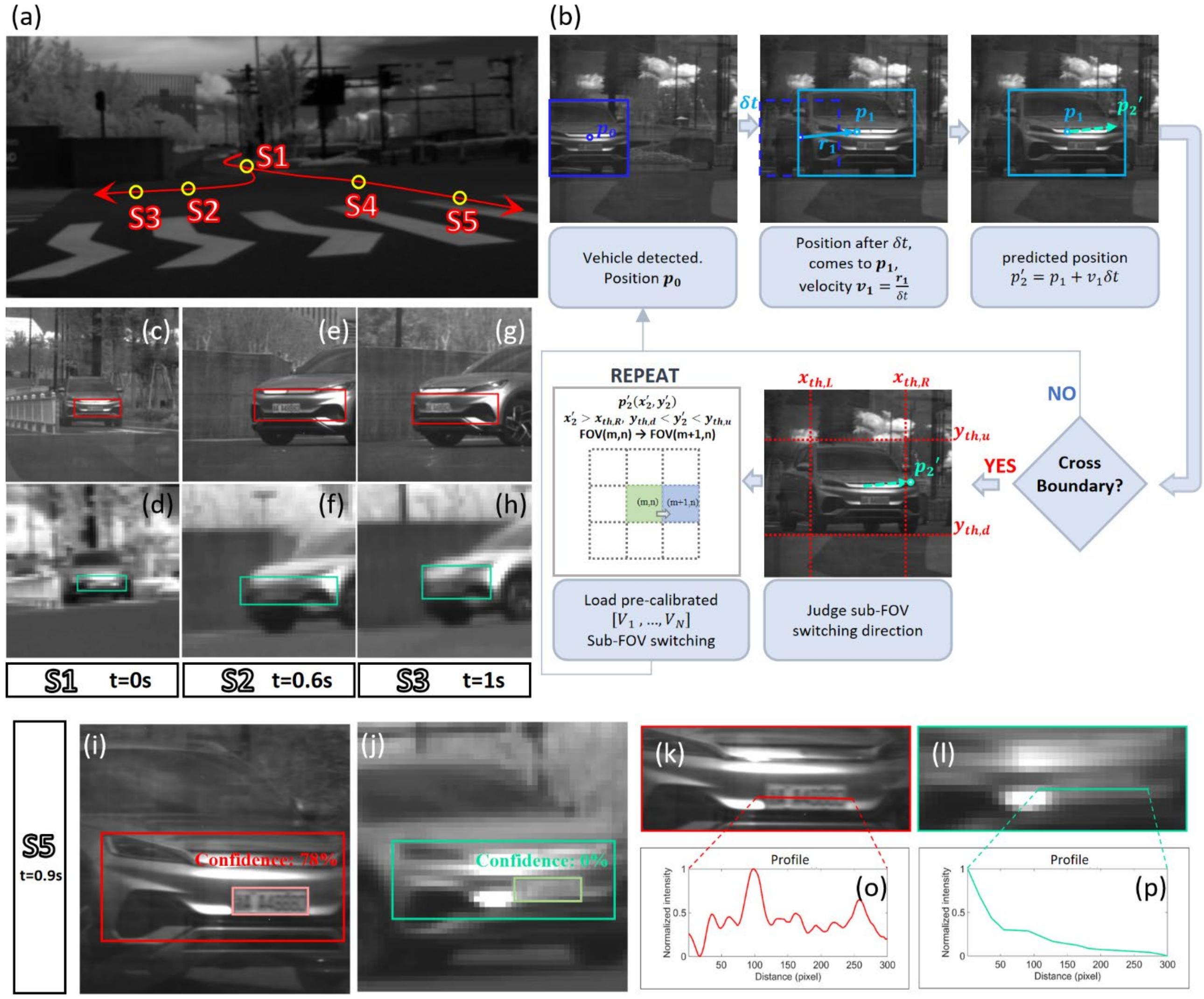

**Fig. 5 | High-speed tracking and high-resolution imaging of moving vehicles and comparison with a commercial wide-FOV camera.** (a) Panoramic view of the intersection captured by the commercial wide-FOV camera from a fixed viewpoint. The red curves and arrows indicate representative vehicle trajectories and traveling directions, and the yellow markers S1–S5 denote selected sequence positions along the trajectories. (b) Illustration of the target-driven tracking and sub-FOV switching logic. The vehicle position is localized in consecutive frames, and its image-plane velocity is estimated to predict the short-term target position. If the predicted position remains within the predefined switching boundaries, the current sub-FOV is retained; once the predicted position crosses one or two boundaries, the corresponding neighboring sub-FOV is selected and the pre-calibrated LCWP voltage set is loaded. (c, e, g) Vehicle images captured by the CaLiBS system at

sequence positions S1, S2, and S3, respectively. (d, f, h) Corresponding cropped vehicle regions from the commercial wide-FOV camera at the same sequence positions. (i, j) Vehicle images at position S5 captured by the CaLiBS system and cropped from the commercial wide-FOV camera, respectively. (k, l) License-plate recognition results obtained from the CaLiBS image and the corresponding wide-FOV image, yielding confidence scores of 78% and 0%, respectively. (m, n) Magnified vehicle-front and license-plate-region views from the CaLiBS image and the corresponding wide-FOV image at S5, respectively. (o, p) Normalized intensity profiles extracted along the marked line cuts in (m) and (n), respectively.

During vehicle motion, pre-calibrated voltage sets were applied to the LCWPs to switch the CaLiBS system to the corresponding steering states, directing the high-resolution instantaneous FOV toward the current vehicle position. **Fig. 5**(c, e, g) show the vehicle images captured by the CaLiBS system at S1, S2, and S3, respectively, and **Fig. 5**(d, f, h) show the corresponding cropped regions from the commercial wide-FOV camera. The time labels indicate the elapsed time along the trajectory. Compared with the cropped wide-FOV images, the CaLiBS images preserve substantially finer vehicle-front details, including headlamp contours, grille features, and license-plate-region information.

To further evaluate the practical utility of the acquired images for recognition-oriented traffic monitoring, license-plate recognition was performed using the images acquired at S5. As shown in **Fig. 5**(i, j), the CaLiBS image retains a clearer vehicle-front and license-plate region than the corresponding wide-FOV crop. A blur-robust license-plate detection and recognition framework based on an optimized YOLOv8 architecture was employed, with detailed method description provided in **Supplementary Note 9**. The license plate in the CaLiBS image was successfully recognized with a confidence score of 78%, whereas the corresponding wide-FOV image yielded a confidence score of 0%, indicating unsuccessful license-plate recognition, as shown in **Fig. 5**(k, l). This result indicates that the enhanced local image quality provided by the CaLiBS system improves not only visual interpretability but also recognition-oriented imaging capability.

The local feature-resolving capability was further compared using magnified vehicle-front and license-plate-region views at S5. **Fig. 5**(m) and **Fig. 5**(n) present the enlarged views from the CaLiBS image and the corresponding wide-FOV image, respectively. The CaLiBS image resolves key vehicle features more clearly, including the headlamp shape, front-body contour, and license-plate-region structure, whereas the corresponding region in the wide-FOV image remains blurred and difficult to identify. To provide a direct comparison of local feature contrast, normalized line-intensity profiles were extracted from the marked line cuts in **Fig. 5**(m, n), as shown in **Fig. 5**(o, p). The profile obtained from the CaLiBS image exhibits pronounced intensity fluctuations with distinguishable peaks and valleys, whereas the profile extracted from the wide-FOV image is much smoother and shows a rapid loss of local modulation. This line-profile comparison further confirms that CaLiBS preserves higher local contrast and finer spatial information during wide-field target tracking.

In practical urban traffic monitoring, reliable recognition of license-plate information and vehicle-front morphology is essential for vehicle identification and target tracking. These results demonstrate the potential of the CaLiBS imaging system for high-resolution tracking and recognition-oriented imaging of moving objects in wide-field surveillance scenarios. Benefiting from all-solid-state electrically controlled beam steering, the system provides a promising route toward high-resolution observation of distant moving targets without mechanical inertia.

**2.5 Fast switching speed of CaLiBS**

For dynamic wide-field imaging, the switching speed of the cascaded system becomes critical, as it determines both the frame rate of full-field raster scanning and the responsiveness of redirecting the high-resolution instantaneous FOV to target-containing sub-FOVs. The switching dynamics are governed by the LCWP response characteristics, which in turn depend on the LC viscosity $\gamma$, the cell gap $d$, and the driving scheme. Because the phase modulation range is set by the product $\Delta nd$, achieving sufficient modulation depth inherently imposes a trade-off with modulation speed. To address this, we implement an overdrive technique[40]: during state transitions, a temporarily elevated voltage is applied to create a stronger electric torque than the steady-state requirement, thereby accelerating molecular reorientation. This approach reduces the rise time to 0.83 ms and the fall time to 44 ms, achieving up to a 13.4-fold speed improvement over conventional driving schemes. Further details are provided in **Supplementary Note 10**.

# 3 Discussion and conclusion

In summary, we have demonstrated a scalable, all-solid-state CaLiBS imaging platform that decouples wide-field angular coverage from local spatial sampling density by electrically steering distinct angular sub-FOVs into a shared high-resolution imaging channel. This architecture circumvents the inherent resolution penalty of conventional wide-FOV cameras and enables two task-adaptive operating modes: full-field raster scanning for static scenes, and selective sub-FOV switching for tracking moving targets. In full-field reconstruction, sequential scanning and stitching yield a 34.7° × 34.7° FOV with an average $MTF_{50}$ of 0.310 cycles/mm (vs. 0.096 cycles/mm for a commercial wide-FOV camera), achieving an 8.6-fold SBP enhancement on the same sensor. In local tracking mode, the system preserves fine details of vehicle fronts and license plates that are blurred in wide-FOV images, enabling successful plate recognition in dynamic traffic scenarios.

A key enabler of this system is an efficient calibration method for compensating oblique-incidence-induced phase errors in LCWPs. The proposed analytical framework predicts half-wave and full-wave voltages as functions of incident angle with RMS error below 0.1 V across all deflection states, ensuring uniform diffraction efficiency exceeding 60% over the full ±15° scanning range. This calibration strategy is essential for maintaining imaging performance in multi-layer cascaded architectures, where cumulative errors would otherwise degrade system throughput.

The cascaded architecture exhibits inherent scalability: increasing the number of LCWP–LCPBG layers causes the deflection state space to grow exponentially[41], enabling wider angular coverage while preserving local resolution. Unlike mechanical scanning systems, the all-solid-state switching is silent, vibration-free, and free from inertia-related limitations, making the platform well-suited for applications requiring simultaneous high resolution, large aperture, and reconfigurable FOV selection.

Several limitations offer clear paths for future improvement. The diffraction efficiency remains below the theoretical maximum due to imperfect polarization conversion at large incident angles and fabrication tolerances, which can be addressed through optimized LCPBG alignment, improved

LCWP retardation accuracy, and anti-reflection coatings. The current reliance on sequential scanning limits temporal resolution for rapidly changing scenes, and tracking fast-moving targets demands faster LC switching—both of which may be addressed by integrating faster LC materials[42,43] or LC metasurfaces[44]. Additionally, as the number of cascaded layers increases, multi-path interference may require generalized matrix-based unmixing beyond the current pairwise symmetric crosstalk model. With ongoing advances in LC materials, nanofabrication, and computational imaging, the CaLiBS platform holds strong potential to serve as a versatile engine for next-generation wide-field observation across remote sensing, autonomous navigation, biomedical imaging, semiconductor inspection, and intelligent surveillance.

## 4 Methods

**Fabrication of LCWP and LCPBG.**

The core components of the beam steering system are LCWPs and LCPBGs, and their key parameters and fabrication considerations are determined based on the system's target of high efficiency, wide FOV and high imaging quality. The LCPBGs used in this work were fabricated using polarization holography based on photoalignment. In this process, a uniform photoalignment layer was first coated onto a clean glass substrate. The sample was then exposed in a Mach–Zehnder interferometric setup, where two coherent ultraviolet beams with orthogonal circular polarization states were superimposed to generate a spatially varying polarization interference pattern on the photoalignment layer[45]. The two orthogonal circularly polarized beams, corresponding to left-handed and right-handed circular polarizations, are typically produced by passing linearly polarized beams through quarter-wave plates. Their interference forms a linearly polarized optical field whose polarization azimuth continuously rotates across the exposure plane, thereby producing a periodic polarization modulation.

The polarization interference pattern was recorded in the photoalignment layer through polarization-sensitive photoinduced anisotropy, which induces the alignment of azo-dye molecules perpendicular to the local polarization direction. As a result, the patterned photoalignment layer imposes a continuous in-plane rotation of the LC director with a periodic spatial profile, forming the anisotropic structure required for the LCPBG. The grating period can be accurately controlled by adjusting the interference angle between the two beams. After photoalignment patterning, the LC layer was formed on the patterned alignment substrate, with its optical retardation optimized for efficient polarization conversion and high first-order diffraction efficiency at the operating wavelength.

The LCWP was designed to control the polarization state of the incident light and thereby switch the diffraction order of the subsequent LCPBG. Each LCWP was fabricated as a transmissive LC cell with a designed cell gap of 6.8 μm and a clear aperture of 50 mm × 50 mm. Briefly, the glass substrates were sequentially cleaned, dried, and coated with an approximately 50 nm-thick polyimide alignment layer. After thermal curing, the polyimide layer was mechanically rubbed along a direction of 45° with respect to the horizontal axis to define the optical axis of the LCWP. Spacer particles were used to control the cell thickness, and the two substrates were aligned, pressed, and bonded with a UV/thermal dual-curing sealant to form the empty LC cell. A high-birefringence LC mixture, LC-BYE7 with $\Delta n = 0.31$, was then filled into the cell by capillary action, followed by sealing of the filling port. Anti-reflection (AR) coatings were applied to reduce Fresnel reflection losses at the optical interfaces, thereby improving the transmittance of each LCWP and the overall efficiency of the cascaded CaLiBS module.

**Acknowledgements**

This work was supported in part by the National Natural Science Foundation of China (Nos. 62405280 and 62431025), Leading Innovative and Entrepreneur Team Introduction Program of Zhejiang (2024R01001).

**Author contributions**

C.X. Liu and M.W. Tang proposed the original idea. C.X. Liu performed the numerical simulations. Q.D. Wang fabricated the samples. C.X. Liu performed the optical measurements with the help of Y.X. Qu and X.X. Wang. All authors discussed the results. C.X. Liu wrote the manuscript with the help of Y.X. Qu. X. Liu, M.W. Tang and K. Wei supervised the project.

**Competing interests**

The authors declare no competing financial interests.

# Supporting Information

## Large-Aperture All-Solid-State Cascaded Liquid-Crystal Beam Steering for High-Resolution Wide-Field Imaging

Chenxi Liu[1], Yongxiang Qu[1], Xiaoxin Wang[2], Leixin Meng[1,2], Xiaohua Feng[1,2], Qidong Wang[3], Guohao Zhang[1], Xuejun Zhang[1,2], Yubing Han[1], Qing Yang[1,2], Xu Liu[1,2], Mingwei Tang[1,2,*] and Kai Wei[1,2,*]

[1] State Key Laboratory of Extreme Photonics and Instrumentation, Zhejiang Key Laboratory of Autonomous Optoelectronic Perception, College of Optical Science and Engineering, Zhejiang University, Hangzhou 310027, China.

[2] ZJU-Hangzhou Global Scientific and Technological Innovation Center, Zhejiang University, Hangzhou 311215, China.

[3] Changchun Institute of Optics, Fine Mechanics and Physics, Chinese Academy of Sciences, Changchun 130033, China

* Address all correspondence to:

Mingwei Tang (tangmw@zju.edu.cn); Kai Wei(kwei@zju.edu.cn);

Table of contents

**Supplementary Note 1：Schematic of a single cascaded LCWP and LCPG module**

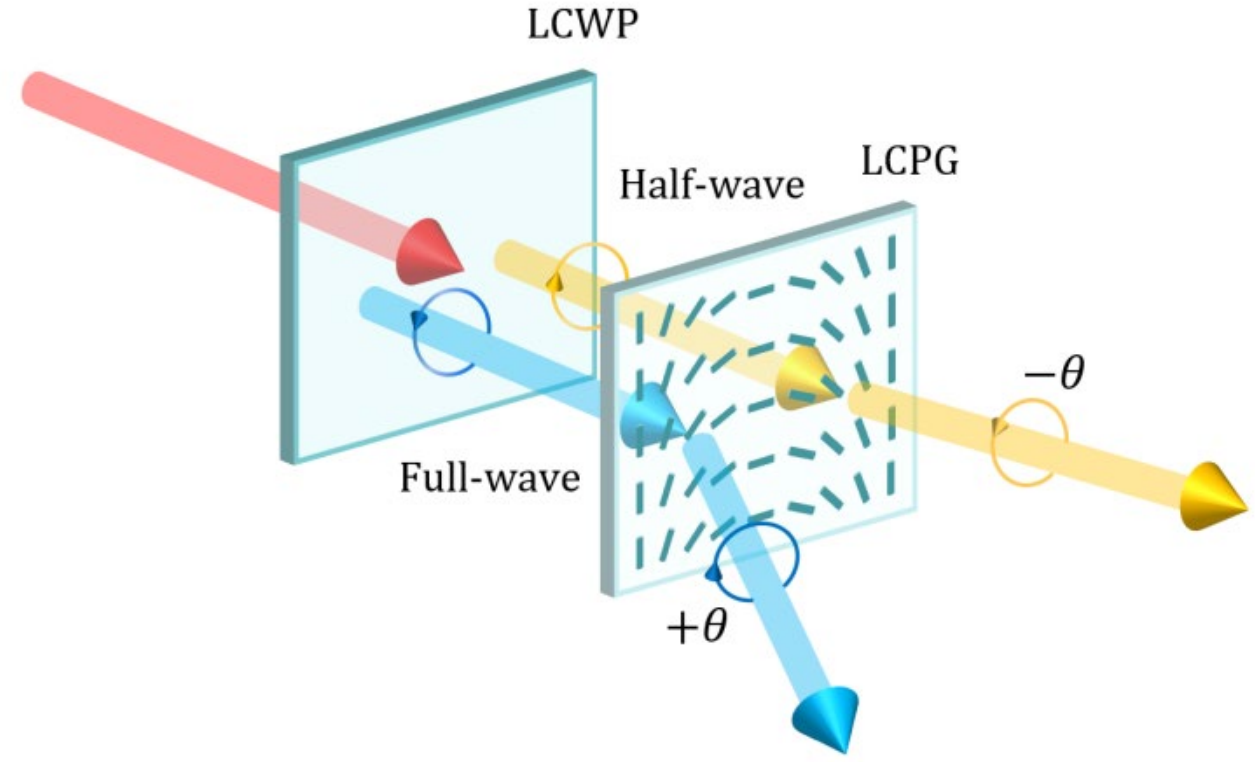

**Fig. S1 |**Schematic of a single steering unit, composed of a LCWP cascaded with a LCPG. The LCWP switches the polarization of incident light, determining whether the LCPG diffracts the beam to the +1st or -1st order.

As shown in **Fig. S1**, the single steering unit provides two switchable beam-deflection states. When the LCWP is operated in the full-wave state, the handedness of the incident circular polarization remains unchanged after passing through the LCWP. The following LCPG then diffracts the beam into one of the first diffraction orders, corresponding to a steering angle of $+\theta$. When the LCWP is switched to the half-wave state, the handedness of the circular polarization is reversed. Since the diffraction direction of the LCPG is determined by the handedness of the incident circular polarization, the same LCPG diffracts the beam into the opposite first diffraction order, corresponding to a steering angle of $-\theta$. Therefore, by electrically switching the retardation state of the LCWP, the output beam can be redirected between the $+\theta$ and $-\theta$ directions without any mechanical movement.

For normal incidence, the deflection angle of the LCPG is determined by its grating period $\Lambda$, which satisfies

$$\boldsymbol{sin\theta} = \frac{\boldsymbol{\lambda}}{\boldsymbol{\Lambda}} \quad (1)$$

where $\lambda$ is the operating wavelength. This binary steering unit serves as the basic element of the cascaded system. By combining multiple units with different grating periods, discrete beam steering over an enlarged field of view can be achieved.

## Supplementary Note 2：Wavefront Characterization of Liquid Crystal Devices

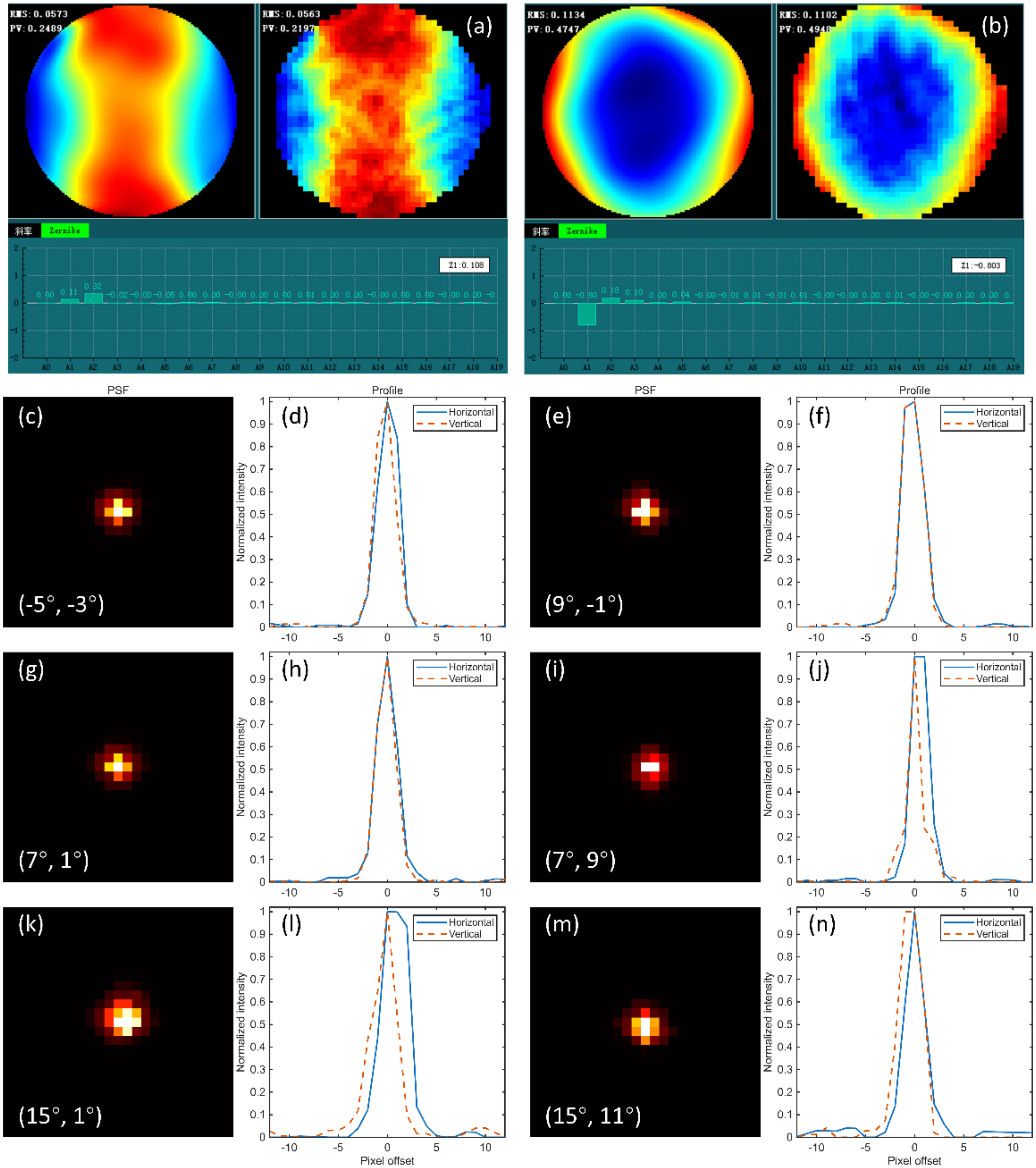

**Fig.S2 |** (a, b). Results of optical wavefront quality of LCWP and LCPG elements respectively measured by a Hartmann–Shack (HS) wavefront sensor. (c-m) Point spread functions (PSFs) and horizonal and vertical profile curves at different incident angle of cascaded LC system switching to corresponding state of FOV.

To ensure minimal wavefront distortion in imaging applications, the optical wavefront quality of the LCWP and LCPG elements was characterized using a Hartmann–Shack (HS) wavefront sensor. The measured spot patterns and reconstructed wavefront maps are shown in **Fig. S2** (a, b).

For the LCWP element, the reconstructed wavefront exhibits a smooth and symmetric profile dominated by low-order aberrations. The root-mean-square (RMS) wavefront error is 0.057λ, with a peak-to-valley (PV) value of 0.249λ. Zernike decomposition indicates that the residual aberration is primarily associated with a weak astigmatic component, while higher-order terms remain negligible. The RMS value well below λ/10 indicates that the LCWP introduces only minimal wavefront distortion and operates close to the diffraction-limited regime.

For the LCPG element, the reconstructed wavefront shows a slightly larger but still well-controlled aberration with an RMS value of 0.113λ and a PV value of 0.475λ. The Zernike analysis reveals that the wavefront is mainly dominated by a defocus-like component, which is expected for spatial phase modulation devices such as liquid-crystal polarization gratings. Higher-order aberrations remain small across the aperture.

Overall, both LCWP and LCPG devices exhibit low wavefront error dominated by low-order aberrations, indicating good optical quality and uniform liquid-crystal alignment. The measured RMS values confirm that the devices introduce negligible wavefront degradation and are suitable for high-quality beam steering and imaging applications.

Meanwhile, the point spread functions (PSFs) of the four-layer cascaded liquid-crystal devices were experimentally characterized, where the incident beams were from different field-of-view (FOV) angles and steered back to the on-axis (0°) direction. As shown in **Fig. S2**(c–n), within the investigated angular range, the measured PSFs at multiple representative FOV positions remained highly concentrated and compact in shape. The focal spots maintained a nearly circular shape with high energy concentration at the center, indicating that the cascaded liquid crystal waveplate–liquid crystal grating system still possesses effective focusing performance for off-axis incident beams.

In addition to the PSF images, the corresponding horizontal and vertical normalized intensity profiles for each field position were also plotted in this paper. For most incident angles, the two profiles almost overlapped, confirming that the focal spots remained approximately rotationally symmetric. At several larger off-axis positions, slight differences in the widths between the horizontal and vertical directions could be observed, along with minor variations in the surrounding low-intensity halo. Nevertheless, no obvious spot splitting, severe stretching, or significant aberration occurred across the entire tested FOV. These results demonstrate that the cascaded liquid-crystal structure can still maintain good wavefront quality over a wide steering range, with PSFs remaining compact at different FOVs, stable focusing performance across the entire FOV, and good aberration control effect.

**Supplementary Note 3：Detailed SBP model of CaLiBS imaging**

In conventional imaging systems, the presence of off-axis aberrations introduces inherent field-dependent behavior, where both the PSF and optical transfer function (OTF) vary as a function of field position. The imaging process can be mathematically formulated as:

$$\mathrm{g}(\mathbf{r}) = \int_{\Omega} o(\boldsymbol{\theta}) h(\boldsymbol{r};\boldsymbol{\theta}) d\boldsymbol{\theta} \tag{2}$$

where $o(\boldsymbol{\theta})$ denotes the object distribution at the field-angle coordinate $\boldsymbol{\theta} = \theta(\theta_x, \theta_y)$, r denotes the image-plane coordinate, and $h(\boldsymbol{r};\boldsymbol{\theta})$ is the field-dependent PSF corresponding to the field point $\boldsymbol{\theta}$. The associated local OTF is given by:

$$H(\boldsymbol{f};\boldsymbol{\theta}) = \mathcal{F}\{h(\boldsymbol{r};\boldsymbol{\theta})\} \tag{3}$$

where $\boldsymbol{f} = (f_x, f_y)$ represents the spatial-frequency coordinate and $\mathcal{F}$ denotes the Fourier transform operator.

For a specified field point $\boldsymbol{\theta}$, the local usable frequency-domain support can be defined as:

$$B(\boldsymbol{\theta}) = \{\boldsymbol{f}: |H(\boldsymbol{f};\boldsymbol{\theta})| \geq \tau\} \tag{4}$$

where $\tau$ is a predetermined threshold parameter that defines effective frequency transfer. Consequently, the effective SBP of a conventional imaging system can be expressed as the integrated accumulation of local transferable frequency-domain support over the entire field:

$$SBP_{conv} \propto \int_{\Omega} |\mathrm{B}(\boldsymbol{\theta})| d\boldsymbol{\theta} \tag{5}$$

Equation (5) reveals that the SBP of a conventional system depends not solely on the geometric FOV, but crucially on the capability of the local OTF to maintain adequate frequency support across the field. As the field angle increases, off-axis aberrations typically intensify, causing $|\mathrm{B}(\boldsymbol{\theta})|$ to diminish substantially at peripheral field positions. Consequently, merely expanding the geometric FOV does not yield a proportional increase in effective SBP.

The proposed CaLiBS overcomes this limitation through a fundamentally different approach. The total field $\Omega$ is partitioned into multiple sub-FOVs $\Omega_m (m = 1,2, \dots, M)$ via the CaLiBS. Each off-axis sub-FOV is sequentially steered into the same high-quality near-axis imaging channel under distinct scanning states. During the m-th scanning state, the image formation can be approximated as:

$$g_m(\boldsymbol{r}) = \int_{\Omega_m} o(\boldsymbol{\theta}) h_0(\boldsymbol{r} - M\boldsymbol{\theta}) d\boldsymbol{\theta} \tag{6}$$

where $h_0(\boldsymbol{r})$ denotes the PSF of the subsequent near-axis imaging channel, and M represents the linear mapping from angular space to the image plane. Equation (6) demonstrates that off-axis sub-FOVs, following deflection by the LC beam steering device, are no longer imaged under their original off-axis PSFs. Instead, they approximately share a common near-axis PSF $h_0$. Accordingly, the local OTF of the m-th sub-FOV can be approximated as:

$$H_{scan}(\boldsymbol{f};\boldsymbol{\theta} \in \Omega_m) \approx H_0(\boldsymbol{f}) \tag{7}$$

where $H_0(\boldsymbol{f}) = \mathcal{F}\{h_0(\boldsymbol{r})\}$ represents the OTF of the near-axis imaging channel. The corresponding local usable frequency-domain support then becomes:

$$B_m \approx B_0 = \{\boldsymbol{f}: |H_0(\boldsymbol{f})| \geq \tau\} \tag{8}$$

This analysis reveals a critical distinction: the CaLiBS system does not expand the frequency-domain support of individual local imaging channels. Rather, through scanning-based angular remapping, it enables multiple sub-FOVs originating from different off-axis positions to be imaged under approximately identical high-quality near-axis conditions, thereby replicating the superior local frequency support B0 across multiple field regions.

The effective SBP of the CaLiBS system can therefore be expressed as:

$$SBP_{scan} \propto \sum_{m=1}^{M} \int_{\Omega_m} |B_0| d\theta = |B_0| \sum_{m=1}^{M} |\Omega_m| \tag{9}$$

When the sub-FOVs are approximately uniform in size, i.e., $|\Omega_{\mathrm{m}}| \approx |\Omega_0|$, this simplifies to:

$$SBP_{scan} \propto M|\Omega_0||B_0| \tag{10}$$

Equation (10) provides fundamental insight into the SBP enhancement mechanism. The observed extension does not originate from direct broadening of the bandwidth of a single OTF. Instead, it results from the sequential reuse and spatial accumulation of the same high-quality near-axis OTF across multiple sub-FOVs. Fundamentally, the system redistributes the field such that the total coverage is expanded while locally transferable bandwidth is preserved to the greatest extent possible, thereby decoupling the conventional trade-off between FOV extension and local resolution degradation.

From an operation perspective, the calibrated voltage lookup table enables two complementary imaging modes. In the full-FOV reconstruction mode, the system sequentially scans a predefined grid of sub-FOVs, followed by crosstalk unmixing and image stitching. In the target-tracking mode, the steering state is selectively updated according to the sub-FOV occupied by a moving object, allowing high-resolution observation of the target without reconstructing the entire panoramic FOV at every time point.

**Supplementary Note 4：Oblique-incidence retardation model and voltage prediction of LCWPs**

For the LCWP, the theoretical operating voltage characteristics can be first analyzed through analytical modeling. In the voltage-free state, LC molecules are uniformly aligned by the alignment layer, with their long molecular axes oriented parallel to the LCWP plane. Upon application of an external electric field via the applied voltage, the LC molecules undergo reorientation driven by the field, achieving a maximum angular rotation approaching 90°. In this saturated state, the long axis of the LC molecules becomes perpendicular to the LCWP plane. Based on this electromechanical response model, the phase modulation profile can be calculated for arbitrary molecular orientations and incident angles:

$$\delta(\theta_i, \theta_d) = \frac{2\pi}{\lambda} \frac{d}{cos\,\theta_t}(n_e(\psi) - n_o) \tag{11}$$

where $\delta(\theta_i, \theta_d)$ represents the phase modulation of the LCWP at an incident angle of $\theta_i$ and the LC molecule tilt angle of $\theta_d$, $\theta_t$ represents the incident angle of light in LC layer, $\psi$ is the angle between the optical axis and the propagation direction in LC layer, $n_e(\psi)$ is the corresponding extraordinary refractive index in the LC layer, and $n_o$ is the ordinary refractive index of LC. To be more specific, $\theta_t$ and $n_e(\psi)$ can be defined as the following equations:

$$n_e(\psi) sin\theta_t = sin\theta_i \tag{12}$$

$$n_e(\psi) = \frac{1}{\sqrt{\frac{cos^2\psi}{n_e^2} + \frac{sin^2\psi}{n_o^2}}} \tag{13}$$

$$cos\psi = \boldsymbol{a} \cdot \boldsymbol{k} \tag{14}$$

Equation (12) represents the refraction formula at the incident interface. Equation (13) gives the calculation formula for the refractive index of extraordinary light in uniaxial crystals. Equation (14) describes the formula for calculating the angle between the light propagation direction and the orientation of LC molecules, where $\boldsymbol{k}$ is the propagation direction vector in LC layer and $\boldsymbol{a}$ is the direction of LC molecules.

Meanwhile, we introduce the mapping between LC molecules rotating angle and driving voltage[1]:

$$\theta_d = \theta' + \frac{\pi}{2} - 2\,arctan\left(e^{-\frac{V-V_C}{V_O}}\right) \tag{15}$$

where $\theta'$ is the pretilt angle of LC molecules, V is the applied voltage, $V_C$ is the threshold voltage, $V_O$ is a constant.

When left-handed and right-handed circularly polarized light is incident on the LCPBG, the diffraction efficiency reaches its maximum value, approaching 100% theoretically. Upon exiting the grating, the polarization state of the light remains circularly polarized with a polarization ellipticity of 45°. Consequently, the polarization state of the light maintains circular polarization, while the chirality may be controlled by the LCWP. When the chirality of the light is reversed, a phase modulation of π from the LCWP is required, and the voltage applied to the LCWP is termed the 'half-wave voltage'; conversely, when the chirality remains unchanged, the LCWP provides a phase modulation of 2π, corresponding to the 'full-wave voltage'. The theoretical half-wave and full-wave voltages can be calculated by combining Equations (11) and (15). The simulated and measured results are shown in **Fig. 2**(a-d). Consequently, the theoretical diffraction efficiency of a cascaded LCWP-LCPBG unit can be expressed as:

$$\eta = T_{LCWP} \times T_{LCPG} \times \eta_{LCPG\pm1} = \begin{cases} T_{LCWP} \times T_{LCPG} \times sin^2\left(\frac{\delta(\theta_i, \theta_d)}{2}\right) & (+1\ order) \\ T_{LCWP} \times T_{LCPG} \times cos^2\left(\frac{\delta(\theta_i, \theta_d)}{2}\right) & (-1\ order) \end{cases} \tag{16}$$

where $T_{LCWP}$ and $T_{LCPG}$ represent the transmission of LCWPs and LCPBGs, $\eta$ is the efficiency of a single cascaded device, $\eta_{LCPG\pm1}$ is the diffraction efficiency of LCPBGs at $\pm1$ order respectively.

## Supplementary Note 5：Conventional calibration workflow of the cascaded LC system

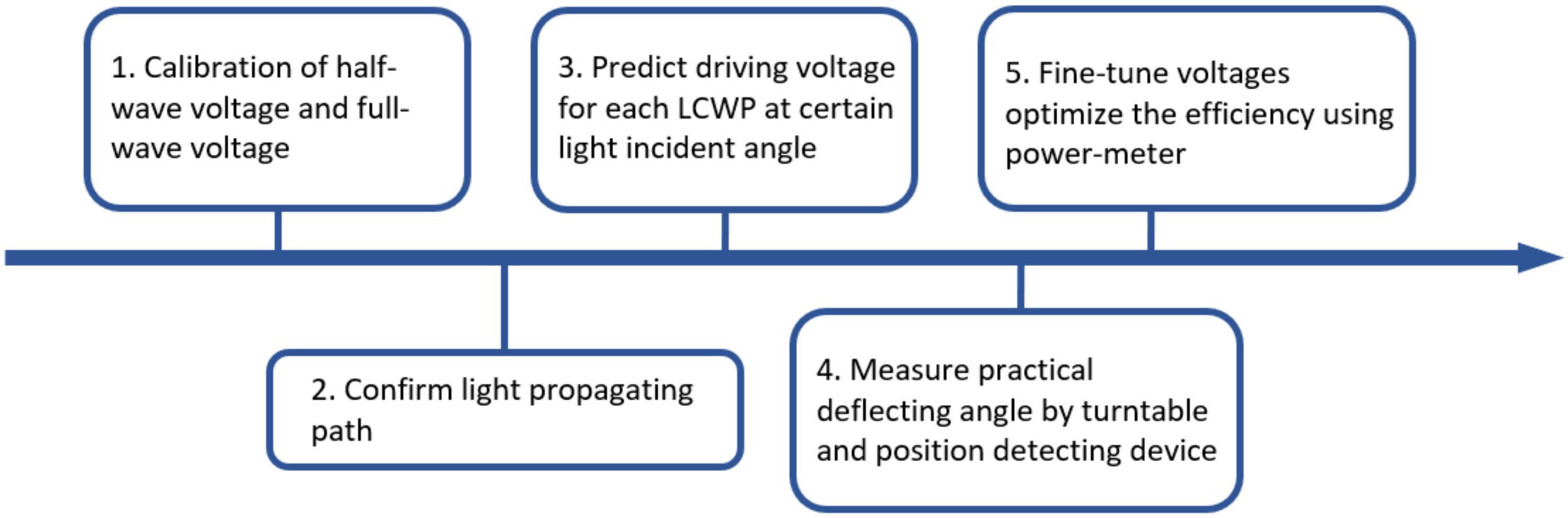

**Fig.S3 |** Main steps of traditional calibration process: 1. Single calibration of LCWP of half-wave and full-wave voltage; 2. Confirmation of light propagation path; 3. Prediction of driving voltage for each LCWP at each state; 4. Deflecting angle measurement by turntable and detector; 5. Efficiency optimization by fine-tuning voltages with power meter.

The overall calibration workflow for the cascaded liquid crystal system is illustrated in **Fig. S3**. The process is structured in five sequential steps, designed to systematically optimize the system performance:

1. Calibration of half-wave and full-wave voltages for individual LCWPs. First, the electro-optic response of each liquid crystal wave plate (LCWP) is calibrated under oblique incidence. This involves measuring the half-wave voltage and full-wave voltage as functions of the incident angle in both the longitude and latitude directions. This data set forms the basis for predicting the driving voltages required for precise polarization control.

2. Confirmation of the light propagation path. Based on the desired output deflection angle, the propagation path of the light beam through the entire cascaded system is determined. This step involves calculating the incident angle on each LCWP and the corresponding diffraction order of each LCPG.

3. Prediction of driving voltages for each LCWP at specific incident angles. Using the single-device calibration data (from Step 1) and the calculated light path (from Step 2), the theoretical driving voltage for each LCWP is predicted. The predicted voltage serves as initial state of the system, ensuring the beam is directed to the ideal position by controlling the polarization state incident on each LCPG.

4. Measurement of the practical deflection angle using a turntable and position-sensing device. The actual deflection angle of the beam is measured using a two-axis turntable and a position-sensing camera. The turntable is rotated to align the system, and the centroid of the output beam is analyzed to determine the real deflection angle, accounting for any manufacturing or assembly errors.

5. Fine-tuning of voltages to optimize efficiency using a power meter. The system's output efficiency is optimized. Starting from the predicted voltages as initial values, a systematic search is performed to find the voltage combination that maximizes the output power, as measured by a power meter. This step yields the optimal driving voltages for the highest efficiency at the target deflection angle.

This integrated calibration procedure ensures that the cascaded LC system operates with both accurate beam deflection and maximum energy efficiency, even under oblique illumination conditions.

**Supplementary Note 6：Calibration of 4-layers CaLiBS System**

**Table S2. Relations between deflecting angles, diffraction orders and incident polarization for 4-layer system**

| Angle Deflection (°) | Diffraction Order/Incident Polarization | | | | | | | |
|---|---|---|---|---|---|---|---|---|
| | Layer (8°) | | Layer (4°) | | Layer (2°) | | Layer (1°) | |
| -15 | -1 | LCP | -1 | LCP | -1 | LCP | -1 | LCP |
| -13 | -1 | LCP | -1 | LCP | -1 | LCP | 1 | RCP |
| -11 | -1 | LCP | -1 | LCP | 1 | RCP | -1 | LCP |
| -9 | -1 | LCP | -1 | LCP | 1 | RCP | 1 | RCP |
| -7 | -1 | LCP | 1 | RCP | -1 | LCP | -1 | LCP |
| -5 | -1 | LCP | 1 | RCP | -1 | LCP | 1 | RCP |
| -3 | -1 | LCP | 1 | RCP | 1 | RCP | -1 | LCP |
| -1 | -1 | LCP | 1 | RCP | 1 | RCP | 1 | RCP |
| 1 | 1 | RCP | -1 | LCP | -1 | LCP | -1 | LCP |
| 3 | 1 | RCP | -1 | LCP | -1 | LCP | 1 | RCP |
| 5 | 1 | RCP | -1 | LCP | 1 | RCP | -1 | LCP |
| 7 | 1 | RCP | -1 | LCP | 1 | RCP | 1 | RCP |
| 9 | 1 | RCP | 1 | RCP | -1 | LCP | -1 | LCP |
| 11 | 1 | RCP | 1 | RCP | -1 | LCP | 1 | RCP |
| 13 | 1 | RCP | 1 | RCP | 1 | RCP | -1 | LCP |
| 15 | 1 | RCP | 1 | RCP | 1 | RCP | 1 | RCP |

For the 4-layer system, the beam deflection angle, the diffraction order of the LCPG in the x or y direction for each layer, and the required incident polarization state for each LCPG are listed in Table S2. According to the required incident polarization state of each LCPG and the output polarization state from the previous LCPG, the operating state (half-wave or full-wave) of the liquid crystal wave plate can be determined to concentrate the incident light energy into the target diffraction order.

The predicted initial driving voltages of the system are calculated based on the wave plate calibration, the determined beam propagation path, and the wave plate operating states obtained from Table S2. The discrepancy between these predicted voltages and the actual optimal driving voltages after system optimization is characterized by the root-mean-square error (RMSE), defined as:

$$\mathrm{RMSE} = \sqrt{\frac{1}{n}\Sigma_{i=1}^{n}\left(V_{i,pred} - V_{i,opt}\right)^2} \quad (17)$$

where n=8 is the number of voltage channels, $V_{i,pred}$ is the predicted initial voltage of the i-th channel, and $V_{i,opt}$ is the optimized optimal voltage. For the scanning matrix with 16×16=256 points, the optical transmission efficiency is controlled by the 8-channel driving voltages at each point. The RMSE between the predicted initial voltage set and the calibrated optimal voltage set is computed for each point, and its distribution is shown in **Fig. S5**. The results demonstrate that the voltage RMSE of all points is below 0.2 V, and more than 91% of the points exhibit an RMSE below 0.1 V. This indicates that the predicted initial voltages provide highly accurate estimates for driving the system toward high-efficiency beam steering.

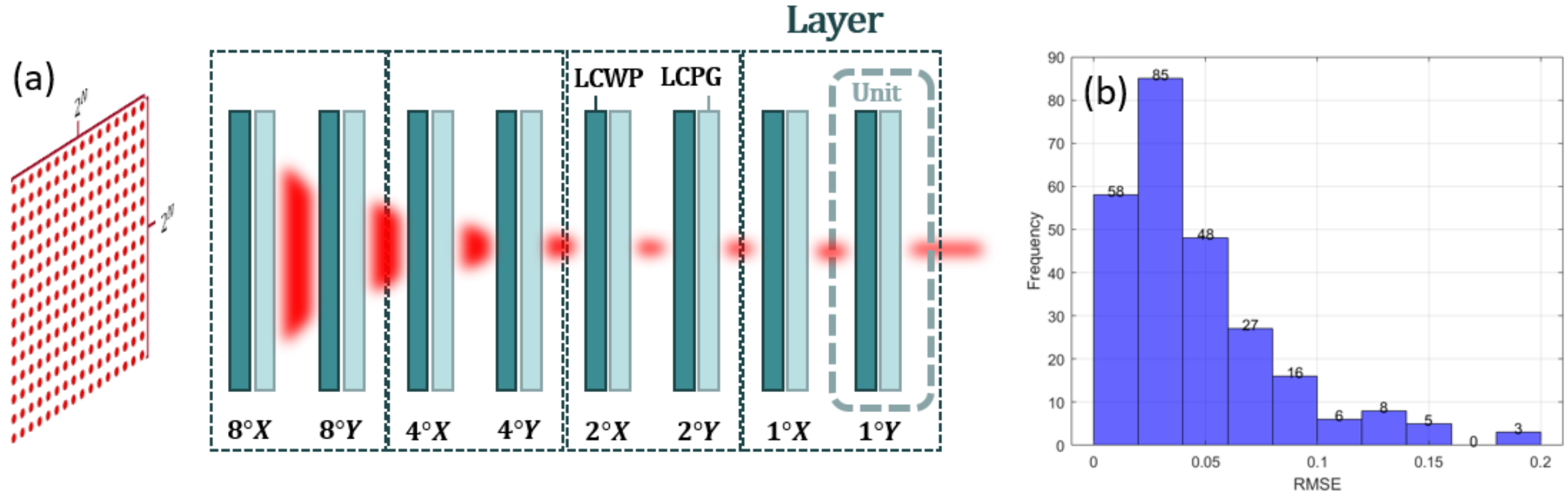

**Fig.S5 |** (a) Schematic diagram of the 4-layer structure receiving a 16×16 dot matrix. (b). RMSE distribution of the 256 scanning points for the 4-layer system. Each RMSE value represents the root-mean-square error between the optimized driving voltages of the eight channels after efficiency optimization and the predicted initial voltages for each scanning point.

## Supplementary Note 7: The Effect of Driving Voltage on Efficiency

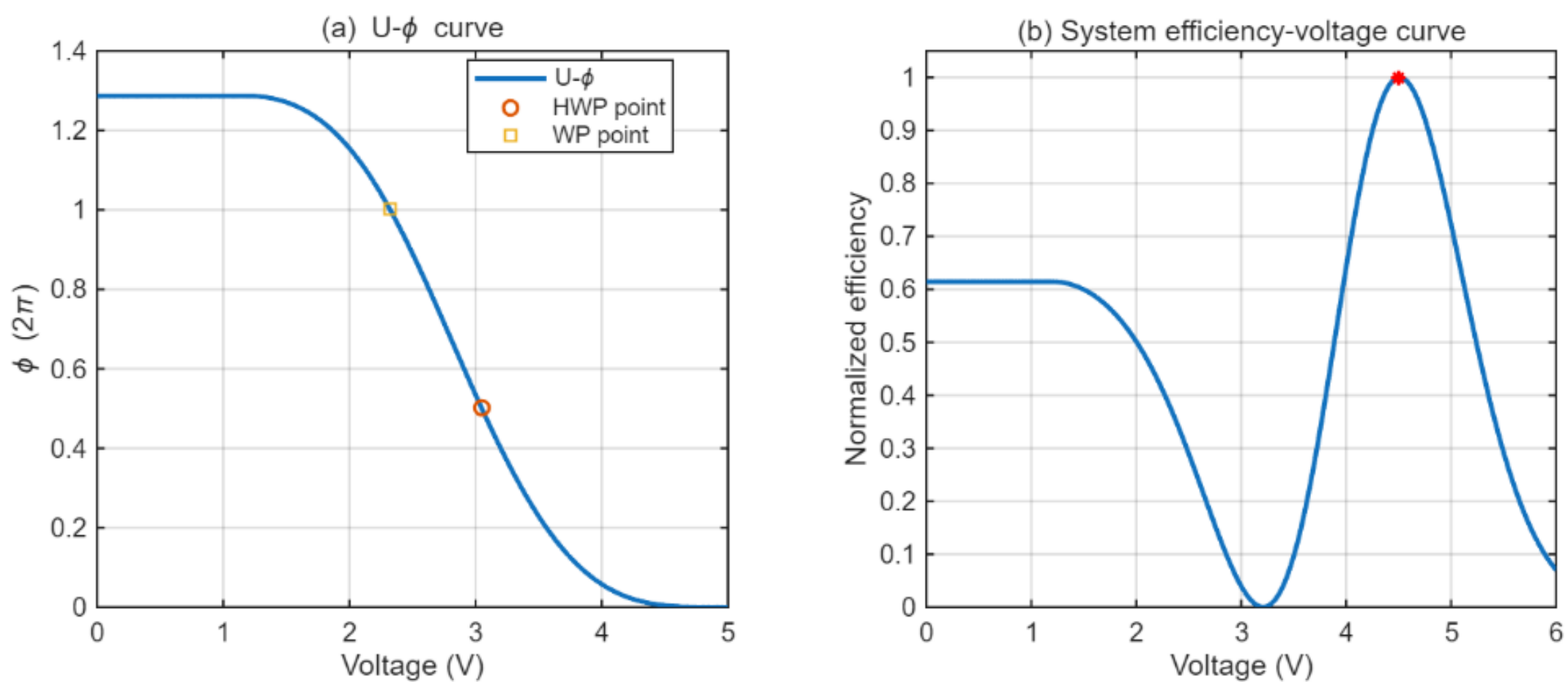

**Fig.S6 |** (a). Mapping of phase retardation ϕ of single LCWP and driving voltages V. (b). Curve of system efficiency as a function of voltage for a single LCWP.

Since effects of the LCWPs on the final efficiency of the cascaded system are mutually independent, each LCWP only affects the efficiency of the cascaded unit where it resides and ultimately contributes to the total system efficiency. Thus, individual analysis of each LCWP is allowed. First, we obtained the equivalent phase retardation of the LCWP under different driving voltages to characterize the $U - \phi$ response of the device (**Fig. S6** (a)). This response reflects the physical process by which the applied voltage modulates the tilt angle of liquid crystal molecules, thereby changing the effective birefringence and phase retardation. Furthermore, given the polarization selectivity of the subsequent LCPG for incident light, the phase retardation of the LCWP not only determines the wavefront modulation state but also governs the conversion ratio between circularly polarized components. Under ideal conditions, the polarization conversion efficiency can be described by $sin^2(\frac{\delta}{2})$. Based on the above analysis, the relationship between the normalized efficiency of the cascaded unit and the driving voltage can be plotted and presented in **Fig. S6** (b).

It is worth noting that the normalized efficiency-voltage curve shown in **Fig. S6** (b) does not represent the actual functional relationship. Combined with the inherent optical properties of practical LCWPs, it can be known that the differences in liquid crystal molecule alignment uniformity, thickness deviation of different LCWP devices, as well as the incident angle of the incident light beam, will all cause a certain degree of translation, stretching or compression changes in the $U - \phi$ response curve of the device. However, it should be emphasized that, limited by the core physical properties of LCWPs, the overall phase coverage range (over 0~2π) and the overall variation trend of their $U - \phi$ response curves (see in **Fig. S6** (a), with the increase of driving voltage, the phase retardation changes monotonically and tends to saturation) remain basically stable, and the morphology of the curve between normalized efficiency and driving voltage derived therefrom will not change essentially. This characteristic provides an important premise for us to construct a peak search and efficiency optimization strategy that is applicable to all LCWP devices and can be widely promoted.

**Supplementary Note 8：Crosstalk**

During the imaging process, we observed that crosstalk tends to occur when the system scans sub-images at diagonal positions of the field of view. Here we analyze the origin of this crosstalk by analyzing the system. For the 4-layer cascaded system, the formula governing the propagation of light incident at the center of a specific field of view can be derived. This formula describes how the beam is deflected layer by layer and finally exits at 0° toward the camera:

$$\begin{bmatrix}\theta_x\\ \theta_y\end{bmatrix} = \begin{bmatrix}S_{1x}\times 8^\circ\\ S_{1y}\times 8^\circ\end{bmatrix} + \begin{bmatrix}S_{2x}\times 4^\circ\\ S_{2y}\times 4^\circ\end{bmatrix} + \begin{bmatrix}S_{3x}\times 2^\circ\\ S_{3y}\times 2^\circ\end{bmatrix} + \begin{bmatrix}S_{4x}\times 1^\circ\\ S_{4y}\times 1^\circ\end{bmatrix} \tag{18}$$

$$\begin{aligned}\begin{bmatrix}\theta'_x\\ \theta'_y\end{bmatrix} = \begin{bmatrix}-\theta_x\\ -\theta_y\end{bmatrix} &= \begin{bmatrix}S'_{1x}\times 8^\circ\\ S'_{1y}\times 8^\circ\end{bmatrix} + \begin{bmatrix}S'_{2x}\times 4^\circ\\ S'_{2y}\times 4^\circ\end{bmatrix} + \begin{bmatrix}S'_{3x}\times 2^\circ\\ S'_{3y}\times 2^\circ\end{bmatrix} + \begin{bmatrix}S'_{4x}\times 1^\circ\\ S'_{4y}\times 1^\circ\end{bmatrix}\\ &= \begin{bmatrix}-S_{1x}\times 8^\circ\\ -S_{1y}\times 8^\circ\end{bmatrix} + \begin{bmatrix}-S_{2x}\times 4^\circ\\ -S_{2y}\times 4^\circ\end{bmatrix} + \begin{bmatrix}-S_{3x}\times 2^\circ\\ -S_{3y}\times 2^\circ\end{bmatrix} + \begin{bmatrix}-S_{4x}\times 1^\circ\\ -S_{4y}\times 1^\circ\end{bmatrix}\end{aligned} \tag{19}$$

where $\theta_x$, $\theta_y$, $\theta'_x$ and $\theta'_y$ represent incident angle, $\mathrm{S_{nx}}$, $\mathrm{S_{ny}}$, $\mathrm{S'_{nx}}$ and $\mathrm{S'_{ny}}$ are diffraction order. It is clear that light from diagonal direction have completely inverse deflection direction at each layer. Then, the state of each LCWP can be confirmed as:

$$\mathrm{S_{LCWPn}} = \begin{cases}\dfrac{S_{n+1x}}{S_{ny}} & (mod(n,2)=0)\\ \dfrac{S_{ny}}{S_{nx}} & (mod(n,2)=1)\end{cases} \tag{20}$$

$\mathrm{S_{LCWPn}}$ is the working state of the n-th LCWP, where a value of 1 represents a full-wave plate and a value of -1 represents a half-wave plate. As for light from diagonal positions of the field of view, to pass through the cascaded system with high efficiency, the state of all 8 LCWPs can be written respectively as:

$$\begin{aligned}\mathrm{S_{LCWP}} &= (\mathrm{S_{LCWP1}}, \mathrm{S_{LCWP2}}, \mathrm{S_{LCWP3}}, \mathrm{S_{LCWP4}}, \mathrm{S_{LCWP5}}, \mathrm{S_{LCWP6}}, \mathrm{S_{LCWP7}}, \mathrm{S_{LCWP8}})\\ &= \left(\frac{\mathrm{S_{1x}}}{S_0}, \frac{\mathrm{S_{1y}}}{\mathrm{S_{1x}}}, \frac{\mathrm{S_{2x}}}{\mathrm{S_{1y}}}, \frac{\mathrm{S_{2y}}}{\mathrm{S_{2x}}}, \frac{\mathrm{S_{3x}}}{\mathrm{S_{2y}}}, \frac{\mathrm{S_{3y}}}{\mathrm{S_{3x}}}, \frac{\mathrm{S_{4x}}}{\mathrm{S_{3y}}}, \frac{\mathrm{S_{4y}}}{\mathrm{S_{4x}}}\right)\end{aligned} \tag{21}$$

$$\begin{aligned}\mathrm{S'_{LCWP}} &= (\mathrm{S'_{LCWP1}}, \mathrm{S'_{LCWP2}}, \mathrm{S'_{LCWP3}}, \mathrm{S'_{LCWP4}}, \mathrm{S'_{LCWP5}}, \mathrm{S'_{LCWP6}}, \mathrm{S'_{LCWP7}}, \mathrm{S'_{LCWP8}})\\ &= \left(\frac{\mathrm{S'_{1x}}}{S_0}, \frac{\mathrm{S'_{1y}}}{\mathrm{S'_{1x}}}, \frac{\mathrm{S'_{2x}}}{\mathrm{S'_{1y}}}, \frac{\mathrm{S'_{2y}}}{\mathrm{S'_{2x}}}, \frac{\mathrm{S'_{3x}}}{\mathrm{S'_{2y}}}, \frac{\mathrm{S'_{3y}}}{\mathrm{S'_{3x}}}, \frac{\mathrm{S'_{4x}}}{\mathrm{S'_{3y}}}, \frac{\mathrm{S'_{4y}}}{\mathrm{S'_{4x}}}\right)\\ &= \left(-\frac{\mathrm{S_{1x}}}{S_0}, \frac{\mathrm{S_{1y}}}{\mathrm{S_{1x}}}, \frac{\mathrm{S_{2x}}}{\mathrm{S_{1y}}}, \frac{\mathrm{S_{2y}}}{\mathrm{S_{2x}}}, \frac{\mathrm{S_{3x}}}{\mathrm{S_{2y}}}, \frac{\mathrm{S_{3y}}}{\mathrm{S_{3x}}}, \frac{\mathrm{S_{4x}}}{\mathrm{S_{3y}}}, \frac{\mathrm{S_{4y}}}{\mathrm{S_{4x}}}\right)\end{aligned} \tag{22}$$

where $\mathrm{S_0}$ represents the initial incident polarization of the light. By analyzing equation (20) and (21), it can be seen that for light incident at diagonal positions, only the state of the first wave plate in the cascaded system differs, while the operating states of all other wave plates remain identical. Since the first wave plate operates under large oblique incidence angles, its polarization conversion efficiency may not reach the ideal 100%. Consequently, diagonally incident light can still pass through the first LCWP and LCPG, and eventually reach the camera via the subsequent LCWPs and LCPGs, leading to crosstalk.

## Supplementary Note 9：YOLOv8-based license-plate detection for blurred vehicle images

In practical intelligent-transportation and security-surveillance scenarios, captured images frequently suffer from severe motion blur and image-quality degradation caused by high vehicle speeds, adverse weather conditions, camera defocus, and other factors. The resulting loss of high-frequency texture information can lead to missed detections and localization errors in conventional object-detection algorithms that rely heavily on distinct edges and local textures.

To address this issue, we developed a blur-robust license-plate detection algorithm based on an optimized YOLOv8 architecture. The proposed network improves the underlying feature-extraction and bounding-box-regression processes in an end-to-end manner. Given an input image, the backbone first generates multiscale feature maps through hierarchical feature extraction. Because local license-plate textures become unreliable under severe blur, the network uses large receptive fields to implicitly learn global vehicle-rear structures and contextual semantic information as prior cues.

During the path-aggregation stage in the neck, spatial- and channel-attention mechanisms are introduced to generate adaptive weighting masks from the high-dimensional feature matrices. These masks suppress complex background interference and direct the network toward potential license-plate regions of interest.

The principal modification is introduced into the regression branch of the prediction head. To address the substantial uncertainty associated with blurred license-plate boundaries, Distribution Focal Loss (DFL) is employed. Conventional bounding-box regression generally treats each target boundary as a fixed and precisely defined coordinate. By contrast, DFL models each boundary coordinate as a probability distribution over neighboring discrete locations and estimates the final coordinate from the expectation of this distribution. Combined with an intersection-over-union-based loss, this uncertainty-aware regression strategy improves robustness under severe blur and boundary attenuation. It therefore produces reliable bounding boxes for subsequent license-plate character recognition.

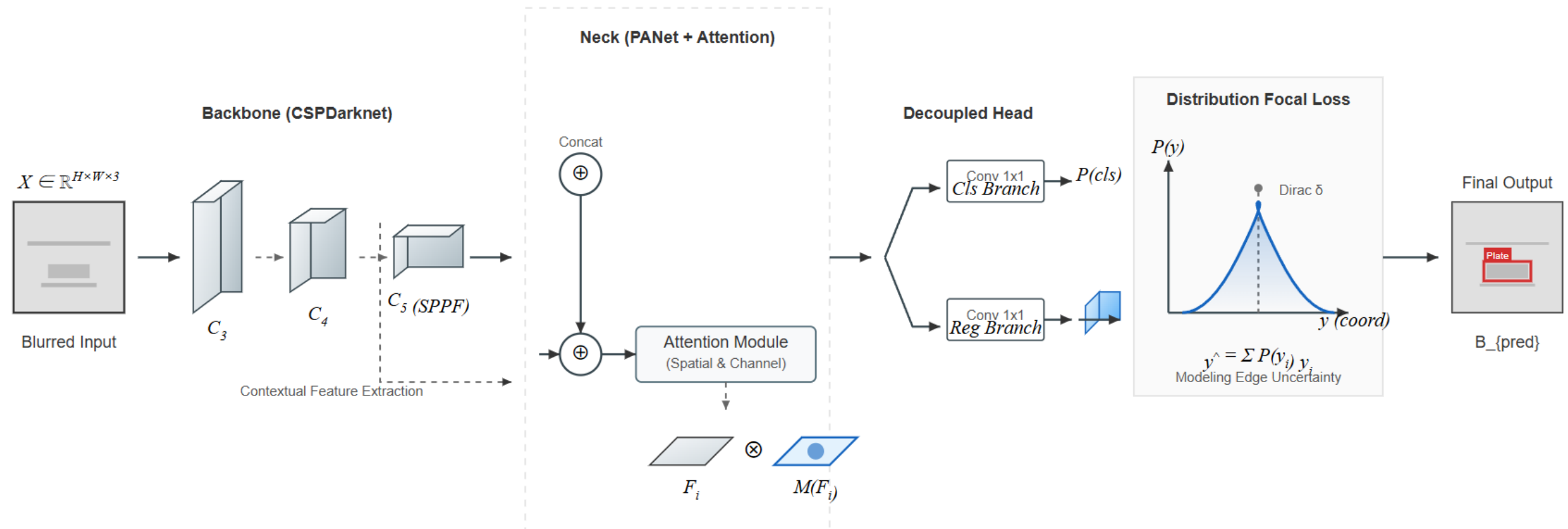

**Fig | S7. Architecture of the proposed YOLOv8-based network for blurred license-plate detection.**

As illustrated in **Fig. S7**, the proposed network consists of five main stages: blurred-image input, contextual feature extraction in the backbone, attention-guided feature fusion in the neck, a decoupled prediction head, and probability-distribution-based boundary regression using DFL. The detailed data flow and underlying mechanisms are described below.

**1. Contextual Feature Extraction in the Backbone**

The network first receives a highly blurred license-plate image X as input. Conventional edge-detection operators are prone to failure when high-frequency texture information is substantially degraded. Therefore, as the input passes through the CSPDarknet-based backbone, the receptive field progressively increases from the shallow feature level $\boldsymbol{C_3}$ to the deeper $\boldsymbol{C_4}$ and $\boldsymbol{C_5}$ levels, with spatial pyramid pooling incorporated at the deepest stage.

Instead of relying exclusively on license-plate characters or sharp plate boundaries, the deeper feature maps capture macroscopic vehicle topology, including the rear-body contour and the relative positions of the taillights. These global structural features serve as contextual semantic priors, allowing the network to infer the probable physical location of the license plate even when its local texture is severely degraded.

**2. Multiscale Feature Fusion and Attention-Based Focusing**

To prevent weak blurred-target features from being attenuated or lost in the deeper layers, the neck adopts a multiscale path-aggregation network, or PANet. As shown in the central part of **Fig. S7**, feature maps extracted at different depths are fused through concatenation operations, denoted by $\oplus$.

An attention module is subsequently embedded in the feature-propagation pathway. The module receives the initially fused feature matrix $\boldsymbol{F_i}$ and calculates its spatial and channel dependencies to generate a weighting-mask matrix $\boldsymbol{M(F_i)}$. The weighting mask is then combined with the original feature map through element-wise tensor multiplication, denoted by $\otimes$. This operation adaptively enhances features associated with potential license-plate regions while suppressing irrelevant background responses. Consequently, the network allocates greater computational attention to the probable license-plate region of interest.

**3. Decoupled Prediction and Probability-Distribution-Based Regression**

The attention-enhanced feature maps are subsequently fed into a decoupled prediction head. Parallel 1×1 convolutional branches are used to separate classification from bounding-box regression, thereby reducing task interference caused by ambiguous blurred features.

For bounding-box regression, the network employs Distribution Focal Loss, as illustrated on the right side of **Fig. S7**. Conventional object detectors generally assume that a target boundary is distinct and uniquely determined, which can be represented conceptually by a Dirac delta distribution. For blurred license plates, however, the transition between the plate boundary and the background is gradual, and the precise boundary location is inherently uncertain.

Instead of directly predicting a single coordinate, DFL estimates the probability $\boldsymbol{P(y_i)}$ that a boundary lies at each neighboring discrete location $\boldsymbol{y_i}$. The final boundary coordinate is obtained from the expectation of the predicted distribution. This probability-distribution-based formulation explicitly models localization uncertainty and produces a more reasonable boundary estimate when the plate edges are weak, diffused, or partially indistinguishable from the background.

**4. Detection Output**

By explicitly modeling boundary uncertainty, the proposed regression strategy enables the network to infer reliable license-plate bounding boxes even under severe motion blur or defocus. The final predicted bounding box, denoted by $\boldsymbol{B_{pred}}$, can therefore maintain relatively high detection confidence and localization accuracy despite substantial edge degradation.

The resulting license-plate region provides a reliable input for the downstream character-recognition module.

## Supplementary Note 10：Speeding up LCWP using overdrive method

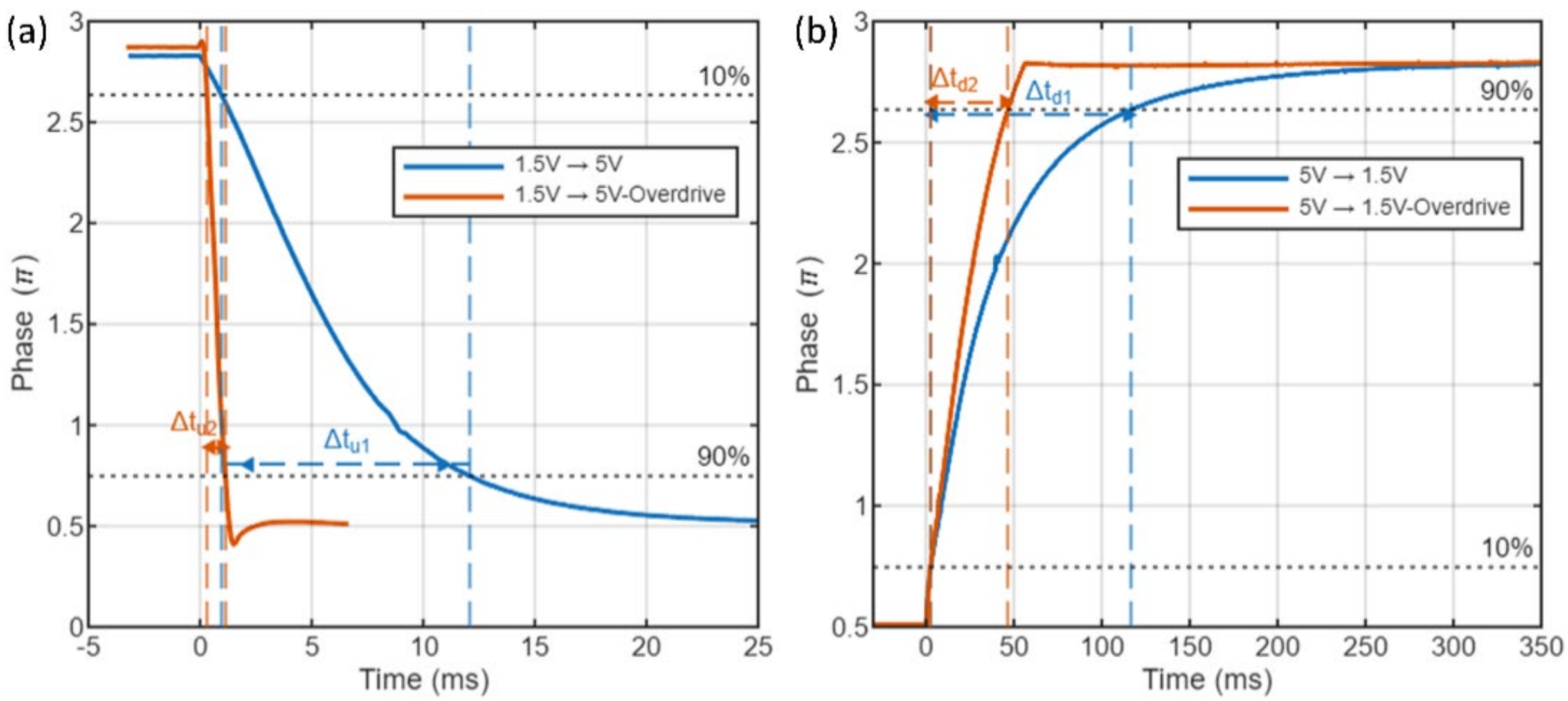

**Fig.S8 |** (a) Phase temporal variation curves of liquid crystal during the rising switching from low driving voltage to high driving voltage. The blue curve represents the conventional driving mode, with the 10%–90% phase rise time defined as $t_{u1}$; the orange curve corresponds to the over-voltage driving mode, and its 10%–90% phase rise time is denoted as $t_{u2}$. (b) Phase temporal variation curves of liquid crystal during the falling switching from high driving voltage to low driving voltage. The blue curve represents the conventional driving mode, with the 10%–90% phase fall time defined as $t_{d1}$; the orange curve corresponds to the over-voltage driving mode, and its 10%–90% phase fall time is denoted as $t_{d2}$.

The dynamic response characteristics of liquid crystal devices mainly depend on the selection of liquid crystal materials and the design of device driving technology. The response time of traditional liquid crystal devices is usually at the millisecond level, which severely restricts their performance in high-speed application scenarios. To break through this technical bottleneck, researchers have developed various response time optimization technologies, among which the over-voltage driving technology has become one of the most practical solutions due to its simple principle[2,3], easy engineering implementation and significant optimization effect.

The basic working mechanism of the over-voltage driving technology is as follows: when liquid crystal molecules need to overcome the resistance torque and elastic torque to complete the orientation adjustment, a driving voltage higher than the target voltage is applied to make the liquid crystal molecules obtain a larger electric field torque, thereby overcoming the obstacles of the resistance torque and elastic torque, and quickly reaching the stable orientation state corresponding to the target voltage. In practical applications, this technology is specifically manifested as: at the moment when the pixel needs to switch the liquid crystal deflection angle, an instantaneous pulse voltage higher than the target voltage is applied to forcefully accelerate the deflection rate of the liquid crystal molecules; when the pixel is close to the target state, the driving voltage drops back to the standard voltage, thereby greatly shortening the time consumption of gray scale switching.

In this experiment, the switching process between two driving voltages (1.5V and 5V) was selected to characterize the switching speed of the cascaded system. The reason is that the switching of these two groups of driving voltages can fully cover the phase change range of more than 2π; under this condition, the liquid crystal molecules need to go through the maximum rotation angle when switching between the two states, and their response speed is the slowest compared with the switching between other phase states, which can more comprehensively and accurately reflect the system switching performance.

In the experiment, by adding polarizers with orthogonal optical axes (horizontal and vertical directions) before and after the liquid crystal wave plate (LCWP) placed with its optical axis tilted at 45°, the phase change of the LCWP was converted into energy change; then the energy change curve was recorded by an oscilloscope to indirectly characterize the variation law of the LCWP phase with time. The measured and extracted phase-time variation curves are shown in **Fig. S8**, where (a) corresponds to the voltage rising process from 1.5V to 5V: the liquid crystal molecules are electrically accelerated and deflected under the action of high electric field torque from the stable state driven by low voltage, and finally reach the stable state driven by high voltage. In **Fig. S8**(a), the blue curve is the phase response curve when the driving voltage is directly switched from 1.5V to 5V, and the orange curve is the phase response curve of the over-voltage driving process where the driving voltage is temporarily driven by 15V from 1.5V and then drops back to 5V. The response time is defined as the time interval for the phase change from 10% to 90%, where the response time of the conventional driving mode is $t_{u1} = 11.12\text{ms}$, and the response time of the over-voltage driving mode is $t_{u2} = 0.83\text{ms}$. The response speed of the over-voltage driving is about 13.4 times higher than that of the conventional driving.

**Fig. S8**(b) corresponds to the voltage falling process from 5V to 1.5V: the liquid crystal molecules gradually fall back to the stable state driven by low voltage after the high voltage is removed and switched to low voltage. At this time, the driving voltage becomes the resistance torque for the movement of liquid crystal molecules; therefore, 0V over-voltage driving is adopted to reduce the resistance during the molecular falling process and accelerate the deflection process of liquid crystal molecules. In **Fig.** S8(b), the blue curve is the phase response curve when the driving voltage is directly switched from 5V to 1.5V, and the orange curve is the phase response curve of the over-voltage driving process where the driving voltage is temporarily driven by 0V from 5V and then switched to 1.5V. Among them, the response time of the conventional driving mode is $t_{d1} = 113.94\text{ms}$, and the response time of the over-voltage driving mode is $t_{d2} = 44\text{ms}$. The response speed of the over-voltage driving is about 2.6 times higher than that of the conventional driving.

**Supplementary References**